\documentclass[12pt]{article}
\usepackage[margin=1.9cm]{geometry}
\usepackage{graphicx}
\usepackage{cite}
\usepackage{xcolor}

\newcommand{\mysection}{\setcounter{equation}{0}\section}

\def\beq{\begin{equation}}
\def\eeq{\end{equation}}
\def\beqa{\begin{eqnarray}}
\def\eeqa{\end{eqnarray}}
 
\begin{document}

\begin{center}
{\Large \bf SMEFT chromomagnetic dipole operator contributions to \\ $t{\bar t}$ production at approximate NNLO in QCD }
\end{center}

\vspace{2mm}

\begin{center}
{\large Nikolaos Kidonakis and Alberto Tonero}\\

\vspace{2mm}

{\it Department of Physics, Kennesaw State University, \\
Kennesaw, GA 30144, USA}

\end{center}

\begin{abstract}
We study higher-order QCD corrections for the production of a top-antitop quark pair ($t{\bar t}$ production) at the Large Hadron Collider in the Standard Model Effective Field Theory. We consider effects of new physics parametrized by the top-quark chromomagnetic dipole operator. We calculate new physics contributions to the total cross section at approximate NNLO, with second-order soft-gluon corrections added to the complete NLO result, including uncertainties from scale dependence and from parton distributions. We also calculate new physics contributions at approximate NNLO for differential distributions in top-quark transverse momentum.
\end{abstract}

\mysection{Introduction}

The top quark is a particularly interesting sector of the Standard Model (SM)  being the heaviest known fundamental particle and coupling most strongly to the electroweak symmetry breaking sector. Furthermore, the production cross section for the top at the Large Hadron Collider (LHC) is quite large such that precision measurements can be made with present and future data that can potentially have groundbreaking sensitivity to beyond the Standard Model (BSM) physics. 

The LHC has now taken a significant amount of data at the TeV scale, but as yet it has found no evidence for BSM physics. While the searches and measurements performed at the LHC cover a tremendous range of BSM theories, it is not necessarily the case that the space of all possible observable deviations from the SM has been searched for in the data. One way to classify BSM models, where new particles are too heavy to be directly produced at the LHC and therefore small deviations from the SM predictions are expected at low energy, is through the Standard Model Effective Field Theory (SMEFT)~\cite{Weinberg:1979sa,Buchmuller:1985jz,Grzadkowski:2010es} where new physics effects are parametrized by a set of higher-dimensional operators organized in a series expansion with increasing operator dimension.

The classification of effective operators contributing to top quark processes~\cite{AguilarSaavedra:2008zc,Zhang:2010dr,AguilarSaavedra:2010zi} marked the beginning of a significant research activity devoted to finding novel ways to constrain higher dimensional operators involving top quarks using LHC data. In the last decade, several groups have developed fitting frameworks to interpret top-quark measurements at the LHC in the context of SMEFT and derive bounds on dimension-six effective operators~\cite{Buckley:2015nca,Buckley:2015lku,Brown:2019pzx,Hartland:2019bjb,Brivio:2019ius,Ellis:2020unq,Ethier:2021bye,Kassabov:2023hbm}. In order to extract consistent results, the theoretical predictions have to match the precision of the experimental measurements which has increased after LHC Run II. Therefore, these analyses use, whenever possible, total and differential SMEFT cross sections which include NLO QCD corrections. The importance of NLO QCD corrections in SMEFT analyses has been discussed in~\cite{Grober:2015cwa,Zhang:2016omx,Passarino:2016pzb,Englert:2019rga}. Recently, automated NLO QCD corrections in the SMEFT have been incorporated in {\small \sc MadGraph5\_aMC@NLO} framework~\cite{MG5} through the {\small \sc SMEFT@NLO} package~\cite{Degrande:2020evl} which allows to compute NLO corrections in QCD for processes involving the presence of several dimension-six effective operators.

In this paper we want to push forward the program of producing more accurate theoretical predictions for top-quark observables in the SMEFT by focusing on the so-called chromomagnetic dipole operator and its effects on top-quark pair production at the LHC. NLO QCD corrections to this operator were calculated for the first time in~\cite{BuarqueFranzosi:2015jrv} and can be computed in an automated way through the aforementioned {\small \sc SMEFT@NLO} package. The chromomagnetic dipole operator is particularly relevant since it enters in many experimental analyses on EFT constraints in $t\bar t$ production~\cite{CMS:2018adi,CMS:2019zct,ATLAS:2022xfj,ATLAS:2022mlu}; therefore, having a precise determination of the modifications induced by this operator to the rates is crucial to improve the fit sensitivity to new physics.

In this work we consider new physics modifications induced by the SMEFT chromomagnetic dipole operator and compute total cross sections and differential distributions in $t{\bar t}$ at approximate next-to-next-to-leading order (aNNLO) in QCD. These results are derived by adding second-order soft-gluon corrections to the complete NLO calculation which includes QCD corrections. 

Soft-gluon resummation \cite{NKGS1,NKGS2,NKnll,NKsingletop,NK2loop,NKsch,NKtW,NKnnll1,NKtch,NKnnll2,NKan3lo1,NKan3lo2,NKtoprev,NK3loop,FK2020,NKresrev,FK2021,NKNY2022,NKNYtqZ,NKAT,KGT,NKCF} is very relevant for top-quark processes in the SM because the cross section receives large contributions from soft-gluon emission near partonic threshold due to the large top-quark mass. This has been well known for a long time for numerous $2 \to 2$ top-quark processes, including top-antitop pair production \cite{NKGS1,NKGS2,NKnll,NKnnll1,NKnnll2,NKan3lo1,NKan3lo2,NKtoprev,KGT} and single-top production \cite{NKsingletop,NKsch,NKtW,NKtch}. More recently, the resummation formalism has been extended \cite{FK2020} to $2 \to 3$ processes, in particular $tqH$ production \cite{FK2021}, $tq\gamma$ production \cite{NKNY2022}, $tqZ$ production \cite{NKNYtqZ}, $t{\bar t} \gamma$ production \cite{NKAT}, and $t{\bar t}W$ production \cite{NKCF}. In all these processes, the soft-gluon corrections are numerically dominant and account for the overwhelming majority of the complete corrections at NLO and NNLO. 

In this work, we calculate for the first time aNNLO QCD corrections in $t{\bar t}$ in the presence of SMEFT operators (more specifically, the chromomagnetic dipole operator) by including soft-gluon corrections at threshold. The paper is organized as follows. In Section 2 we introduce the SMEFT framework and identify the dimension-six operator relevant for this study. In Section 3, we describe the soft-gluon resummation formalism for $t{\bar t}$ partonic processes. In Section 4, we present numeical results for the total cross section through aNNLO. In Section 5, we present results for the top-quark differential distributions in transverse momentum $p_T$ through aNNLO at 13 and 13.6 TeV LHC energies. We conclude in Section 6.   

\mysection{SMEFT contributions to $t{\bar t}$ production at the LHC}

The language of the Standard Model Effective Field Theory (SMEFT) is very suitable for phenomenological studies in presence of heavy BSM physics. In particular, when new physics degrees of freedom are much heavier than the energy scales relevant to top-quark pair production, one can describe the most general departures from the SM predictions in terms of higher dimension effective operators. The leading contributions come from dimension-six operators~\cite{Grzadkowski:2010es} and can be parametrized in terms of an effective Lagrangian as follows:
\begin{equation}
{\cal L}_{\rm SMEFT}={\cal L}_{\rm SM}+\sum_i \frac{c_i}{\Lambda^2}{\cal O}_{i}+{\rm h.c.}\,,
\end{equation}
where $\Lambda$ represents the scale of new physics and $c_i$ are dimensionless Wilson coefficients.

The complete dimension-six basis consists of 59 operators (modulo all the possible flavor combinations) and their presence can modify top-quark pair production at the LHC. All possible dimension-six operators that  belong to the SMEFT Lagrangian and give rise to top-quark interactions that contribute at tree level to the $t{\bar t}$ process are presented in~\cite{AguilarSaavedra:2018nen}. In this work we focus on the so-called chromomagnetic dipole operator, which is an operator that is frequently considered in recent experimental analyses on SMEFT in $t\bar t$ production~\cite{CMS:2018adi,CMS:2019zct,ATLAS:2022xfj,ATLAS:2022mlu}. The chromomagnetic dipole operator is defined as follows:
\begin{equation}\label{eftop}
\frac{c_{tG}}{\Lambda^2}{\cal{O}}_{tG}+{\rm h.c.}
\end{equation}
where
\beq\label{opdef}
{\cal{O}}_{tG}=g_S \bar q_{3L} \sigma_{\mu\nu}T^A t_R\tilde \varphi G_A^{\mu\nu} \, .
\eeq
Here $g_S$ is the strong coupling, $T^A$ are the $SU(3)$ generators, $\sigma_{\mu\nu}=i[\gamma_{\mu},\gamma_{\nu}]/2$ with $\gamma_{\mu}$ the Dirac matrices, $q_{3L}=(t\quad b)_L$, $\varphi$ is the Higgs doublet, and $G_{\mu\nu}^A$ is the gluon field strength tensor. The $c_{tG}$ coefficient is a dimensionless Wilson coefficient which, in this work, is taken to be real. In general, after electroweak symmetry breaking, the chromomagnetic dipole operator induces anomalous $gtt$ couplings that can be parameterized by the following effective vertex~\cite{AguilarSaavedra:2008zc}
\begin{equation}
\label{attvertex}
V_{g tt}=-g_S\bar t \frac{i\sigma^{\mu\nu}q_\nu}{m_t}
(d_V^g +id_A^g \gamma_5)T_At G^A_\mu \,,
\end{equation}
where $q$ is the photon momentum and $d_V^g$ and $d_A^g$ are anomalous couplings related to the top-quark chromo-magnetic and chromo-electric dipole moment, respectively. The explicit contributions of the coefficients to these anomalous couplings are given by the relations
\begin{equation}\label{anom_couplings}
d_V^g =\sqrt{2}{\rm Re}[c_{tG}]\frac{\upsilon\, m_t}{\Lambda^2} \, , \qquad \qquad d_A^g= \sqrt{2}{\rm Im}[c_{tG}]\frac{\upsilon\, m_t}{\Lambda^2} \, ,
\end{equation}
where $\upsilon=246$ GeV is the Higgs vacuum expectation value (vev).

The top-antitop production process in the presence of the chromomagnetic dipole operator can occur at leading order (LO) through quark-antiquark annihilation and gluon fusion channels. 
We consider SMEFT partonic processes  
\beq
f_{1}(p_1)\, + \, f_{2}\, (p_2) \rightarrow t(p_t) \, + \, {\bar t}(p_{\bar t})  \, ,
\label{processes}
\eeq
where $f_1$ and $f_2$ are quarks or gluons in the protons. At LO, the partonic processes are $q{\bar q} \to t{\bar t}$ and $gg \to t{\bar t}$. We define $s=(p_1+p_2)^2$, $t=(p_1-p_t)^2$, $u=(p_2-p_t)^2$, and $s_4=s+t+u-2m_t^2$ where $m_t$ is the top quark mass. The LO partonic differential cross section can be written as
\beq
\frac{d^2{\hat{\sigma}}^{(0)}_{q{\bar q} \to t\bar t}}{dt \, du}= F^B_{ab \to t{\bar t}}(c_{tG}) \; \delta(s_4)\, ,
\label{LO}
\eeq
where $F^B_{ab \to t{\bar t}}(c_{tG})$ represents the Born matrix element squared divided by $16\pi s^2$, which is a quadratic function of the effective operator coefficient $c_{tG}$. Therefore, this LO differential cross section can also be expressed as a polynomial of second degree in the Wilson coefficient $c_{tG}$. Indeed, for the quark-antiquark annihilation channel we can write
\beq\label{bornqua}
F^B_{q{\bar q} \to t \bar t}(c_{tG})=F^B_{q{\bar q} \to t \bar t,0}+\frac{c_{tG}}{\Lambda^2}F^B_{q{\bar q} \to t \bar t,1}
+\frac{c_{tG}^2}{\Lambda^4} F^B_{q{\bar q} \to t \bar t,2}\,,
\eeq
while for the gluon fusion channel we can write
\beq\label{bornglu}
F^B_{gg \to t \bar t}(c_{tG})=F^B_{gg \to t \bar t,0}+\frac{c_{tG}}{\Lambda^2}F^B_{gg \to t \bar t,1}
+\frac{c_{tG}^2}{\Lambda^4} F^B_{gg\to t \bar t,2}\,.
\eeq
For the quark-antiquark annihilation channel, the SM contribution is
\beq
F^B_{q{\bar q} \to t \bar t,0} = \frac{{4 \alpha_s^2 \pi}}{{9s^4}}\left(4m_t^2s + t^2 + u^2 -2m_t^4 \right) \, ,
\eeq
the SMEFT interference term is
\beq
F^B_{q{\bar q} \to t \bar t,1} = \frac{{16\sqrt{2} \alpha_s^2 \pi m_t \upsilon }}{{9s^4}}
\left[4m_t^2s + \left( t + u \right)^2 -4m_t^4 \right] \, ,
\eeq
and the SMEFT squared term is
\beq
F^B_{q{\bar q} \to t \bar t,2}=\frac{{4 \alpha_s^2 \pi\upsilon^2}}{{9s^3}}\left[ 3s^2+2s(t+u)-(t-u)^2
\right] \, .
\eeq
For the gluon fusion channel, the SM contribution is
\beqa
F^B_{gg \to t \bar t,0}&=&-\frac{{\alpha_s^2 \pi}}{{768s^4\left( m_t^2 - t \right)^2 \left( m_t^2 - u \right)^2}}
\left[7 s^2+9(t-u)^2 \right]
\nonumber \\ && \hspace{-5mm} \times
\left[3 s^4+(t-u)^4 +12 s^3 (t+u)+4 s (t-u)^2 (t+u)+4 s^2 (3t^2+2 t u+3 u^2)\right] \, ,
\eeqa
the SMEFT interference term is
\beq
F^B_{gg \to t \bar t,1} 
=\frac{{\sqrt{2}\alpha_s^2 \pi m_t \upsilon}}{{96s^2\left( m_t^2 - t \right)^2 \left( m_t^2 - u \right)^2}}
\left[7s^4+2s^2(t-u)^2-9(t-u)^4 \right]\, ,
\eeq
and the SMEFT squared term is
\beqa
F^B_{gg \to t \bar t,2} 
&=&\frac{{\alpha_s^2 \pi  \upsilon^2}}{{384s^3\left( m_t^2 - t \right)^2 \left( m_t^2 - u \right)^2}}
\left[53s^6+32s^5(t+u)-65s^4(t-u)^2 \right.
\nonumber \\ && \left.
{}-32 s^3 (t-u)^2(t+u)+3s^2(t-u)^4+9(t-u)^6
\right] \, .
\eeqa
Notice that the LO expressions in the equations above are manifestly symmetric in $t$ and $u$.

Early studies about constraints on anomalous top-quark gauge couplings considered the contributions of the SMEFT chromomagnetic dipole operator at tree level, see for instance~\cite{Tonero:2014jea,Fabbrichesi:2014wva,Barducci:2017ddn}. Then, NLO QCD corrections to the operator of Eq.~(\ref{eftop}) became available and were computed for the first time in~\cite{BuarqueFranzosi:2015jrv}. In this work, we improve the order of the QCD corrections to the chromomagnetic dipole operator in top-quark pair production processes by going to aNNLO.

\mysection{Soft-gluon corrections and resummation for $t{\bar t}$ production}

In this section, we briefly discuss the resummation formalism that we use for the calculation of soft-gluon corrections in $t{\bar t}$ production in the Standard Model \cite{NKGS1,NKGS2,NKnll,NK2loop,NKnnll1,NKnnll2,NKan3lo1,NKan3lo2} and in SMEFT. We note that the overall theoretical formalism for these calculations is the same for the SM and SMEFT. The origin of the soft-gluon corrections is from the emission of soft (low-energy) gluons, which result in partial cancellations of infrared divergences between real-emission and virtual diagrams close to partonic threshold for the production of a $t{\bar t}$ pair (see \cite{NKresrev} for a review).

We consider $t{\bar t}$ production in proton-proton collisions with SM and SMEFT partonic processes  
as in Eq. (\ref{processes}). When considering soft-gluon corrections in single-particle-inclusive kinematics, 
we use the partonic threshold variable $s_4$ introduced below Eq. (\ref{processes}) which may also be reexpressed as $s_4=(p_{\bar t}+p_g)^2-m_t^2$, with $p_g$ the momentum of an additional gluon in the final state. Near partonic threshold $p_g \to 0$ and, thus, $s_4 \to 0$. 
The soft-gluon corrections appear in the perturbative series as logarithms of $s_4$, i.e. $[(\ln^k(s_4/m_t^2))/s_4]_+$, with $0 \le k \le 2n-1$ at $n$th order in $\alpha_s$.
 
The resummation of soft-gluon corrections is a consequence of the factorization properties of the (differential) cross section under Laplace transforms. We first write the differential hadronic cross section, $d\sigma_{pp \to t{\bar t}}$, as a convolution of the differential partonic cross section, $d{\hat \sigma}_{ab \to t{\bar t}}$, with the pdf $\phi_{a/p}$ and $\phi_{b/p}$, as 
\beq
d\sigma_{pp \to t{\bar t}}(c_{tG})=\sum_{a,b} \; 
\int dx_a \, dx_b \,  \phi_{a/p}(x_a, \mu_F) \, \phi_{b/p}(x_b, \mu_F)  \, 
d{\hat \sigma}_{ab \to t{\bar t}}(s_4, \mu_F,c_{tG})  \, ,
\label{factorized}
\eeq
where $c_{tG}$ is the effective operator coefficient, $\mu_F$ is the factorization scale, and $x_a$, $x_b$ are momentum fractions of partons $a$, $b$, respectively, in the colliding protons.

We take Laplace transforms, with transform variable $N$, of the partonic cross section, $d{\tilde{\hat\sigma}}_{ab \to t{\bar t}}(N)=\int_0^s (ds_4/s) \,  e^{-N s_4/s} \, d{\hat\sigma}_{ab \to t{\bar t}}(s_4)$, and of the pdf, ${\tilde \phi}(N)=\int_0^1 e^{-N(1-x)} \phi(x) \, dx$. Then, Eq. (\ref{factorized}) gives the parton-level $N$-space expression 
\beq
d{\tilde \sigma}_{ab \to t{\bar t}}(N,c_{tG})= {\tilde \phi}_{a/a}(N_a, \mu_F) \, {\tilde \phi}_{b/b}(N_b, \mu_F) \, d{\tilde{\hat \sigma}}_{ab \to t{\bar t}}(N, \mu_F,c_{tG}) \, .
\label{factN}
\eeq

A refactorization of the cross section is made as a product of a short-distance hard function $H_{ab \to t{\bar t}}$, a soft function $S_{ab \to t{\bar t}}$ for noncollinear soft-gluon emission, and distributions $\psi_{i/i}$ for collinear gluon emission from the incoming partons:
\beq
d{\tilde{\sigma}}_{ab \to t{\bar t}}(N, c_{tG})={\tilde \psi}_{a/a}(N_a,\mu_F) \, {\tilde \psi}_{b/b}(N_b,\mu_F) \, {\rm tr} \left\{H_{ab \to t{\bar t}} \left(c_{tG}\right) \, {\tilde S}_{ab \to t{\bar t}} \left(\frac{\sqrt{s}}{N \mu_F} \right)\right\} \, ,
\label{refactN}
\eeq
where $H_{ab \to t{\bar t}}$ and $S_{ab \to t{\bar t}}$ are process-dependent matrices in an appropriate color-tensor basis, and a trace is taken in the above equation. We note that the chromomagnetic dipole operator enters only in the hard scattering matrix $H_{ab \to t {\bar t}}$, which depends explicitly on $c_{tG}$, while the other functions in the above equation are the same as in the SM. The hard and soft functions are $2 \times 2$ matrices for $q{\bar q} \to t{\bar t}$ and $3 \times 3$ matrices for $gg \to t{\bar t}$. The renormalization-group evolution of $S_{ab \to t{\bar t}}$ involves a soft anomalous dimension matrix for each partonic channel, resulting in the exponentiation of logarithms of the transform variable conjugate to $s_4$: 
\beq
\mu_R\frac{d{\tilde S_{ab \to t{\bar t}}}}{d\mu_R}=\left(\mu_R \frac{\partial}{\partial \mu_R}
+\beta(g_s, \epsilon)\frac{\partial}{\partial g_s}\right) {\tilde S}_{ab \to t{\bar t}}
=-\Gamma_{\! S \, ab \to t{\bar t}}^{\dagger} \; {\tilde S}_{ab \to t{\bar t}}
-{\tilde S}_{ab \to t{\bar t}} \; \Gamma_{\! S \, ab \to t{\bar t}}
\label{RGE}
\eeq 
where $\mu_R$ is the renormalization scale, $\Gamma_{\! S \, ab \to t{\bar t}}$ is the soft anomalous dimension matrix, $g_s^2=4\pi\alpha_s$, and $\beta(g_s, \epsilon)=-g_s\epsilon/2+\beta(g_s)$ in $4-\epsilon$ dimensions with $\beta(g_s)=\mu_R\, dg_s/d\mu_R$ the QCD beta function. We note that $\Gamma_{\! S \, ab \to t{\bar t}}$ is a $2 \times 2$ matrix and $\Gamma_{\! S \, gg \to t{\bar t}}$ is a $3 \times 3$ matrix.

Comparing Eqs. (\ref{factN}) and (\ref{refactN}), we derive an expression for the partonic cross section 
\beq
d{\tilde{\hat \sigma}}_{ab \to t{\bar t}}(N,\mu_F,c_{tG})=
\frac{{\tilde \psi}_{a/a}(N_a, \mu_F) \, {\tilde \psi}_{b/b}(N_b, \mu_F)}{{\tilde \phi}_{a/a}(N_a, \mu_F) \, {\tilde \phi}_{b/b}(N_b, \mu_F)} \; \,  {\rm tr} \left\{H_{ab \to t{\bar t}}\left(c_{tG}\right) \, {\tilde S}_{ab \to t{\bar t}}\left(\frac{\sqrt{s}}{N \mu_F} \right)\right\} \, .
\label{partonicN}
\eeq

The renormalization-group evolution of ${\tilde S}_{ab \to t{\bar t}}$ as well as the other $N$-dependent functions (${\tilde \psi}$ and ${\tilde \phi}$) in Eq. (\ref{partonicN}) leads to the resummation of the corrections from collinear and soft gluons. The resummed cross section is given by 
\beqa
d{\tilde{\hat \sigma}}_{ab \to t{\bar t}}^{\rm resum}(N,\mu_F,c_{tG}) &=&
\exp\left[\sum_{i=a,b} E_{i}(N_i)\right] \, 
\exp\left[\sum_{i=a,b} 2 \int_{\mu_F}^{\sqrt{s}} \frac{d\mu}{\mu} \gamma_{i/i}(N_i)\right] 
\nonumber \\ && \hspace{-2mm} 
\times {\rm tr} \left\{H_{ab \to t{\bar t}}\left(\alpha_s(\sqrt{s}),c_{tG}\right) {\bar P} \, \exp \left[\int_{\sqrt{s}}^{{\sqrt{s}}/N} \frac{d\mu}{\mu} \, \Gamma_{\! S \, ab \to t{\bar t}}^{\dagger} \left(\alpha_s(\mu)\right)\right] \right.
\nonumber \\ && \hspace{8mm}
\left. 
\times {\tilde S}_{ab \to t{\bar t}} \left(\alpha_s\left(\frac{\sqrt{s}}{N}\right)\right) \;
P\, \exp \left[\int_{\sqrt{s}}^{{\sqrt{s}}/N} \frac{d\mu}{\mu} \, \Gamma_{\! S \, ab \to t{\bar t}} \left(\alpha_s(\mu)\right)\right] \right\} 
\label{resummed}
\eeqa
where $P$ (${\bar P}$) denotes path ordering in the same (reverse) sense as the integration variable $\mu$. The first exponent, $E_i$,  in Eq. (\ref{resummed}) involves the lightlike cusp anomalous dimension and resums universal collinear and soft contributions from the incoming partons, and the second exponential shows the factorization-scale dependence in terms of the anomalous dimension $\gamma_{i/i}$ of the parton density $\phi_{i/i}$. Explicit expressions can be found in the review of Ref. \cite{NKresrev}; for a NNLO calculation of soft-gluon corrections they are needed through two loops. The LO cross section, given in the previous section, involves the product of the hard and soft functions at LO, while the NLO cross section, to which we match, involves them at one-loop. The resummation of noncollinear soft-gluon emission is performed via the soft anomalous dimension matrices $\Gamma_{\! S \, q{\bar q} \to t{\bar t}}$ and $\Gamma_{\! S \, gg \to t{\bar t}}$ following Eq. (\ref{RGE}); these matrices have been calculated fully at one loop \cite{NKGS1,NKGS2} and two loops \cite{NK2loop,FNPY,NKnnll1}, and some pieces are known at three loops \cite{NKresrev} and four loops \cite{NK4loop}.

The inverse transform of Eq.~(\ref{resummed}) does not require any prescription when performed at fixed perturbative order, and it gives, at each perturbative order, the form of the soft-gluon corrections as plus distributions of logarithms of $s_4$, as discussed earlier. Approximate next-to-next-to-next-to-leading-order (aN$^3$LO) corrections to $t{\bar t}$ production were calculated in Refs. \cite{NKan3lo1,NKan3lo2} and, with the additional inclusion of electroweak corrections, in Ref. \cite{KGT}. More details on the formalism and for past applications to $t{\bar t}$ production can be found in Refs. \cite{NKtoprev,NKGS1,NKGS2,NKnll,NK2loop,NKnnll1,NKnnll2,NKan3lo1,NKan3lo2,NKresrev}. 

\mysection{Total cross sections at 13 and 13.6 TeV}

In this section we present results for the total cross section for $t{\bar t}$ production at the LHC, including the effect of the dimension-six operator in Eq.~(\ref{opdef}). We consider diagrams containing at most one effective operator insertion. The cross section can be written as a polynomial of second degree in the Wilson coefficient, 
\begin{equation}\label{sigma_smeft}
\sigma(c_{tG})=\beta_0 + \frac{c_{tG}}{(\Lambda/1 {\rm TeV})^2} \beta_1
+\frac{c_{tG}^2}{(\Lambda/1 {\rm TeV})^4} \beta_2 \, ,
\end{equation}
where $\beta_0$ is the SM cross section, $\beta_1$ is the contribution derived from the interference of SM diagrams and diagrams with one SMEFT insertion, while $\beta_2$ corresponds to the cross section terms derived by squaring diagrams with one SMEFT insertion. Notice that we have normalized the SMEFT coefficients in the equation above in order for the $\beta_i$ to have all the same dimension.

\begin{table}[htbp]
\begin{center}
\begin{tabular}{|c|c|c|c|} \hline
\multicolumn{4}{|c|}{SM and SMEFT contributions to $ t{\bar t}$ cross sections at LHC 13 TeV} \\ \hline
   & $\beta_0$ (pb) & $\beta_1$ (pb) & $\beta_2$ (pb) \\ \hline
LO (LO pdf)  & $575^{+186}_{-132}{}^{+7}_{-7} $& $184^{+59}_{-42}{}^{+3}_{-2} $ & $33.4^{+11.3}_{-7.9}{}^{+0.7}_{-0.5} $ \\ \hline
LO (NLO pdf)  & $488^{+143}_{-104}{}^{+8}_{-8} $& $156^{+45}_{-33}{}^{+2}_{-3} $ & $28.1^{+8.7}_{-6.2}{}^{+0.6}_{-0.4} $ \\ \hline
LO (NNLO pdf)  & $487^{+142}_{-103}{}^{+10}_{-6} $& $155^{+46}_{-32}{}^{+4}_{-2} $ & $28.1^{+8.6}_{-6.1}{}^{+0.7}_{-0.4} $ \\ \hline
NLO (NLO pdf) & $730^{+86}_{-86}{}^{+13}_{-11} $& $233^{+27}_{-27}{}^{+4}_{-4} $ & $41.9^{+4.8}_{-5.0}{}^{+0.8}_{-0.7} $ \\ \hline
NLO (NNLO pdf) & $730^{+85}_{-86}{}^{+14}_{-10} $& $232^{+27}_{-27}{}^{+5}_{-3} $ & $41.8^{+4.8}_{-5.0}{}^{+1.0}_{-0.6} $ \\ \hline
aNNLO (NNLO pdf) & $814^{+28}_{-46}{}^{+16}_{-11} $& $259^{+9}_{-15}{}^{+6}_{-3} $ & $46.6^{+1.6}_{-2.6}{}^{+1.1}_{-0.7} $ \\ \hline
\end{tabular}
\caption[]{SM and SMEFT contributions to the $t{\bar t}$ cross section (in pb) at LO, NLO, and aNNLO, with scale and pdf uncertainties, at the LHC with $\sqrt{S}=13$ TeV, $m_t=172.5$ GeV, and MSHT20 pdf.}
\label{table1}
\end{center}
\end{table}

\begin{table}[htbp]
\begin{center}
\begin{tabular}{|c|c|c|c|} \hline
\multicolumn{4}{|c|}{SM and SMEFT contributions to $t{\bar t}$ cross sections at LHC 13.6 TeV} \\ \hline
 & $\beta_0$ (pb)& $\beta_1$ (pb)& $\beta_2$ (pb)\\ \hline
LO (LO pdf)  &  $638^{+203}_{-145}{}^{+11}_{-8} $& $204^{+65}_{-46}{}^{+4}_{-3} $ & $37.3^{+12.4}_{-8.8}{}^{+0.7}_{-0.6} $ \\ \hline
LO  (NLO pdf)  &  $540^{+156}_{-114}{}^{+9}_{-8} $& $172^{+50}_{-36}{}^{+3}_{-2} $ & $31.4^{+9.5}_{-6.9}{}^{+0.6}_{-0.6} $ \\ \hline
LO (NNLO pdf)  &  $540^{+155}_{-113}{}^{+10}_{-7} $& $172^{+49}_{-36}{}^{+3}_{-2} $  & $31.3^{+9.4}_{-6.8}{}^{+0.7}_{-0.5} $ \\ \hline
NLO (NLO pdf) & $810^{+95}_{-95}{}^{+14}_{-12} $& $258^{+30}_{-30}{}^{+4}_{-4} $ & $46.7^{+5.4}_{-5.6}{}^{+0.9}_{-0.8} $ \\ \hline
NLO (NNLO pdf) & $809^{+94}_{-94}{}^{+16}_{-11} $& $257^{+29}_{-30}{}^{+5}_{-3} $ & $46.6^{+5.3}_{-5.5}{}^{+1.0}_{-0.7} $ \\ \hline
aNNLO (NNLO pdf) & $902^{+31}_{-50}{}^{+18}_{-12} $& $287^{+10}_{-16}{}^{+6}_{-3} $ & $52.0^{+1.8}_{-2.9}{}^{+1.1}_{-0.8} $ \\ \hline
\end{tabular}
\caption[]{SM and SMEFT contributions to the $t{\bar t}$ cross section (in pb) at LO, NLO, and aNNLO, with scale and pdf uncertainties, at the LHC with $\sqrt{S}=13.6$ TeV, $m_t=172.5$ GeV, and MSHT20 pdf.}
\label{table2}
\end{center}
\end{table}

The complete LO and NLO QCD results are calculated using {\small \sc MadGraph5\_aMC@NLO} \cite{MG5} and the {\small \sc SMEFT@NLO} model \cite{Degrande:2020evl}. We use MSHT20~\cite{MSHT20} pdf in our calculations and we set the factorization and renormalization scales equal to each other, with this common scale denoted by $\mu$. For the total cross section, the central results are obtained by setting $\mu=m_t$, where $m_t$ is the top-quark mass and it is taken to be $m_t$ = 172.5 GeV. Scale uncertainties are obtained by varying the common scale $\mu$ in the range $m_t/2\leq \mu\leq 2 m_t$, and pdf uncertainties are also computed.

In Table 1 and Table 2 we show the SM and SMEFT contributions to the total rates, with scale and pdf uncertainties, for $t{\bar t}$ production at 13 TeV and 13.6 TeV center-of-mass energy, respectively, at LO, NLO, and aNNLO using MSHT20 pdf at various orders. In the tables, we use the exact NNLO QCD result for the SM value \cite{Czakon:2013goa,Catani:2019iny}, which is well known and is, in any case, very well approximated by the aNNLO calculation, as shown in~\cite{NKtoprev}. 

The scale uncertainties are similar for SM and SMEFT contributions and improve by going to aNNLO: they are roughly $+12$\%  $-12\%$ at NLO and around $+3.4\%$ $-5.5\%$ at aNNLO for both LHC energies. In addition to scale variation, there are also uncertainties from the pdf which are much smaller than the scale uncertainties.

By inspecting Table 1 and Table 2 one can compute individual $K$-factors for each $\beta_i$ term; we observe that both the NLO over LO and the aNNLO over LO $K$-factors of the SMEFT terms ($\beta_1$ and $\beta_2$) are very similar (the difference is less than 1\%) to the corresponding $K$-factors of the SM term ($\beta_0$)~\cite{NKnnll1,NKan3lo1}. This shows that the $K$-factor similarity between SM and SMEFT contributions, which was first presented at NLO in~\cite{BuarqueFranzosi:2015jrv}, holds also at aNNLO. For instance, considering MSHT20 NNLO pdf at 13 TeV center of mass energy, we find that the NLO over LO $K$-factors are given by
\begin{equation}\label{kf13tot}
\frac{\beta_0^{\rm NLO}}{\beta_0^{\rm LO}}=1.50 \, , \qquad\qquad
\frac{\beta_1^{\rm NLO}}{\beta_1^{\rm LO}}=1.50 \, , \qquad\qquad
\frac{\beta_2^{\rm NLO}}{\beta_2^{\rm LO}}=1.49 \, ,
\end{equation}
while the aNNLO over LO $K$-factors are 
\begin{equation}\label{kf13.6tot}
\frac{\beta_0^{\rm NNLO}}{\beta_0^{\rm LO}}=1.67 \, , \qquad\qquad
\frac{\beta_1^{\rm aNNLO}}{\beta_1^{\rm LO}}=1.67 \, , \qquad\qquad
\frac{\beta_2^{\rm aNNLO}}{\beta_2^{\rm LO}}=1.66 \, .
\end{equation}
The similarity between SMEFT and SM $K$-factors holds also when considering pdf corresponding to
each order of the perturbative calculation. Moreover, the $K$-factors at 13.6 TeV center-of-mass energy are practically identical to the ones evaluated at 13 TeV, showing a weak dependence on the center-of-mass energy. In general, this relation of the SM and SMEFT $K$-factors does not hold if we change the $\mu$ scale. For instance, we checked that, at 13 TeV, if we set $\mu=m_T=(p_T^2+m_t^2)^{1/2}$, the NLO/LO $K$-factor of the $\beta_1$ term stays very close to the SM (less than 1\% difference) while the $\beta_2$ $K$-factor differs by roughly 2\%.
We would like to stress that the relation between SM and SMEFT $K$-factors is also operator dependent. We checked that, for another SMEFT operator, the four-fermion operator ${\cal{O}}_{tq}^{(8)}$, the SMEFT and SM $K$-factors are very different already at NLO. Moreover, for that operator, the soft-gluon corrections are not dominant and they differ greatly from the SM ones.

\begin{figure}[htbp]
\begin{center}
\includegraphics[width=88mm]{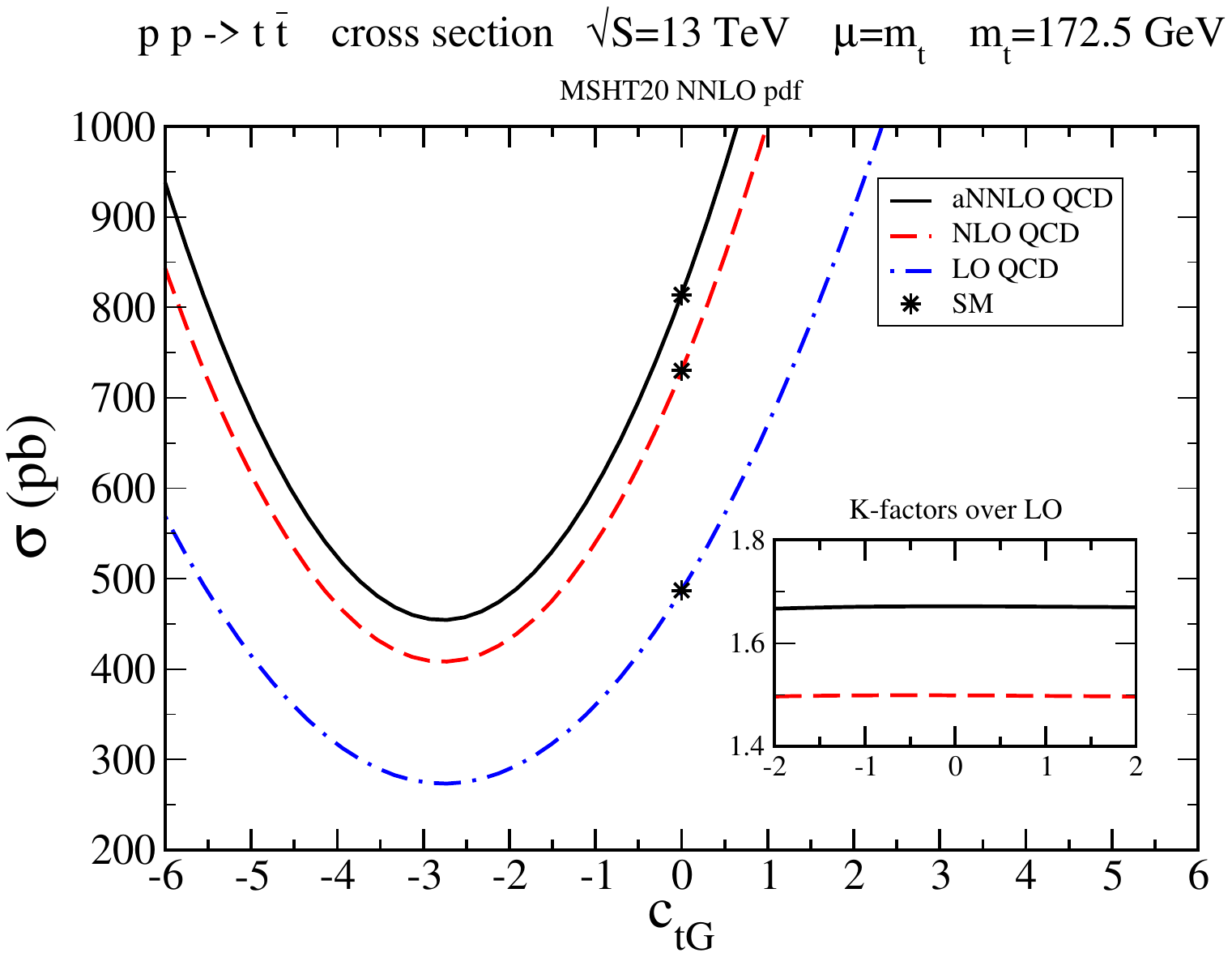}
\includegraphics[width=88mm]{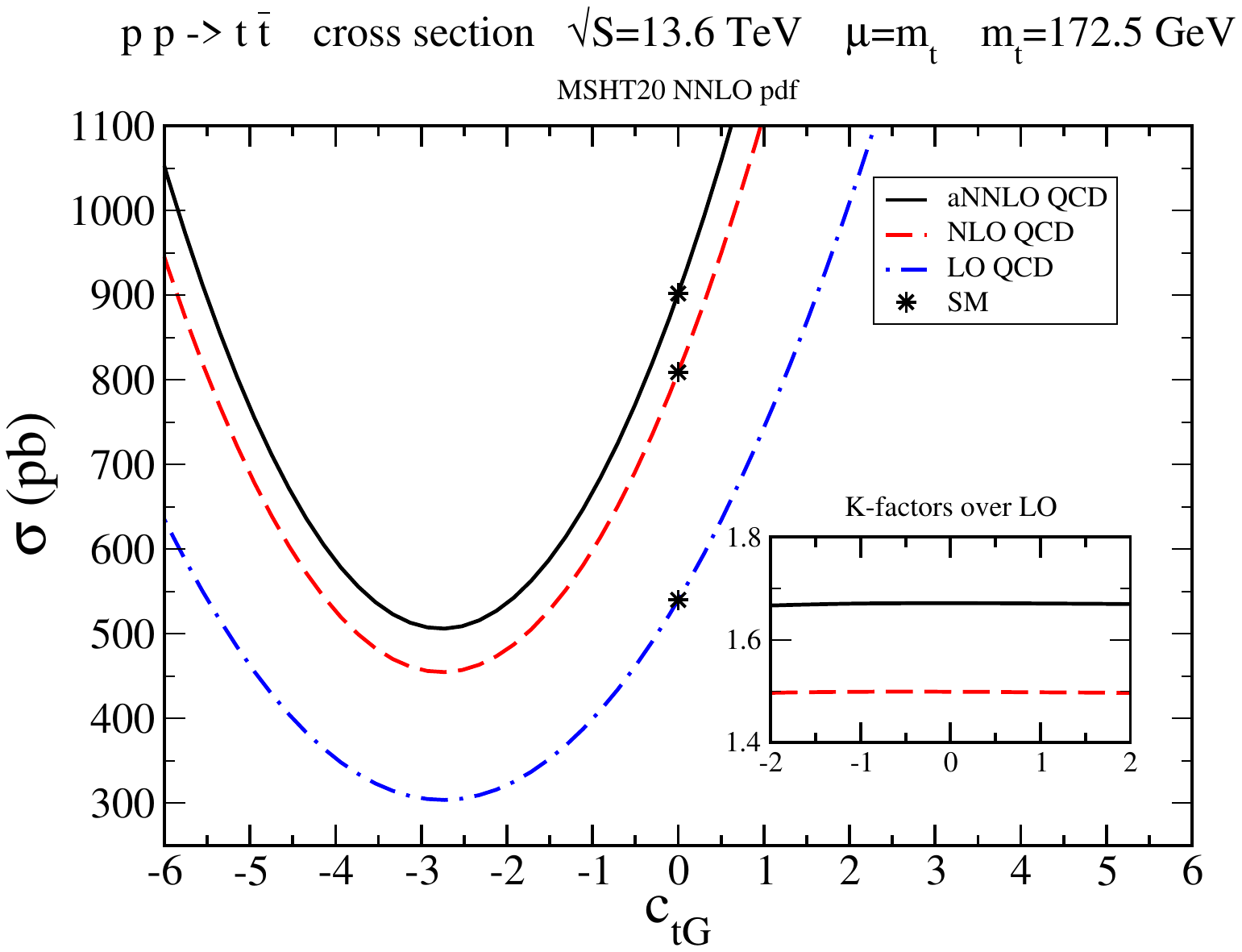}
\caption{Dependence on the effective operator coefficient ($c_{tG}$) of the LO (blue dot-dashed), NLO (red dashed), and aNNLO (black solid) total cross section for $t{\bar t}$ production in $pp$ collisions at LHC 13 TeV (left) and 13.6 TeV (right), using MSHT20 NNLO pdf. The black stars highlight the SM prediction of the cross section. The inset plots show the NLO over LO and aNNLO over LO $K$-factors.}
\label{xseckf}
\end{center}
\end{figure}

In Fig.~\ref{xseckf} we show  the total cross section for  $t{\bar t}$ production in proton-proton collisions as a function of the effective operator coefficient $c_{tG}$ at 13 TeV (left plot) and 13.6 TeV (right plot). The plots show the central value (obtained for $\mu=m_t$) of the LO, NLO and aNNLO cross sections, all three computed using MSHT20 NNLO pdf. These plots are useful in showing how each order in the series contributes to the final aNNLO result. The black stars indicate the SM prediction value of the total cross sections. The inset plots display the $K$-factors of the total cross section relative to LO, i.e. the NLO over LO and aNNLO over LO ratios, 
defined as
\begin{equation}\label{K-fac}
\frac{\sigma^{\rm NLO}(c_{tG})}{\sigma^{\rm LO}(c_{tG})}\qquad\qquad{\rm and} \qquad\qquad
\frac{\sigma^{\rm aNNLO}(c_{tG})}{\sigma^{\rm LO}(c_{tG})}
\end{equation}
where $\sigma(c_{tG})$ is given by Eq.~(\ref{sigma_smeft}). 
From the plots we can see that these $K$-factors' dependence on the effective coupling $c_{tG}$ is flat, as already suggested by the numerical values obtained in Eq.~(\ref{kf13tot}) and Eq.~(\ref{kf13.6tot}).

In order to estimate the effects that our aNNLO QCD calculations might have on a SMEFT global fit analysis, we   compare the theoretical predictions of Table 1 with the most precise measurements of the total cross section of top-quark pair production at 13 TeV and set limits on the size of the effective operator coefficient $c_{tG}$.
Here we consider as experimental input, the most recent ATLAS and CMS results of $829\pm 15$ pb~\cite{ATLAS:2023gsl} and $791 \pm 25$ pb~\cite{CMS:2021vhb}, respectively. In order to find the 95\% CL exclusion limits, we set $\Lambda=1$ TeV and construct the following chi-squared function:
\begin{equation}\label{chi2f}
\chi^2(c_{tG})=\frac{[\sigma_{\rm exp}-\sigma(c_{tG})]^2}{\delta\sigma_{\rm exp}^2+\delta\sigma(c_{tG})^2} \, , 
\end{equation}
where $\sigma_{\rm exp}$ is one of the aforementioned experimental measurements of the total $t\bar t$ cross section, with associated uncertainty $\delta\sigma_{\rm exp}$, and $\sigma(c_{tG})$ is the theoretical prediction of the cross section as function of the effective operator coefficient $c_{tG}$. The theoretical uncertainty of the total cross section $\delta\sigma(c_{tG})$ is computed by firstly symmetrizing the total uncertainties of each contribution $\beta_i$ (which are in turn obtained by combining in quadrature scale and pdf errors). The symmetrization is realized  by taking the most conservative envelope, namely $\delta \bar\beta_i={\rm max}\{\delta \beta_i^{\rm up}, \delta \beta_i^{\rm down}\}$. Then, these symmetric uncertainties are combined to give the total theory uncertainty as $\delta\sigma(c_{tG})=\delta\bar\beta_0 + c_{tG}\delta\bar\beta_1+c_{tG}^2\delta\bar\beta_2 $.

\begin{figure}[htbp]
\begin{center}
\includegraphics[width=88mm]{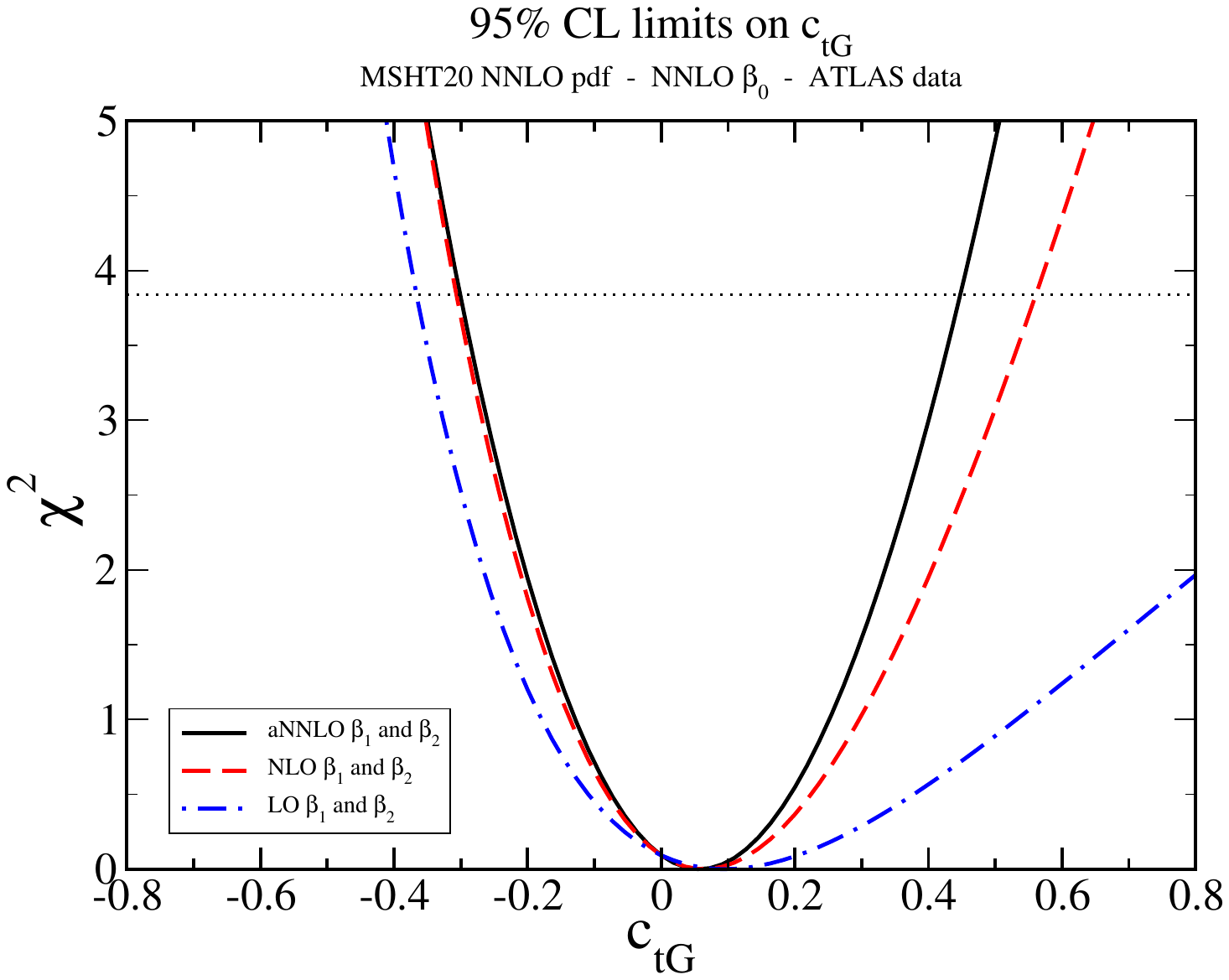}
\includegraphics[width=88mm]{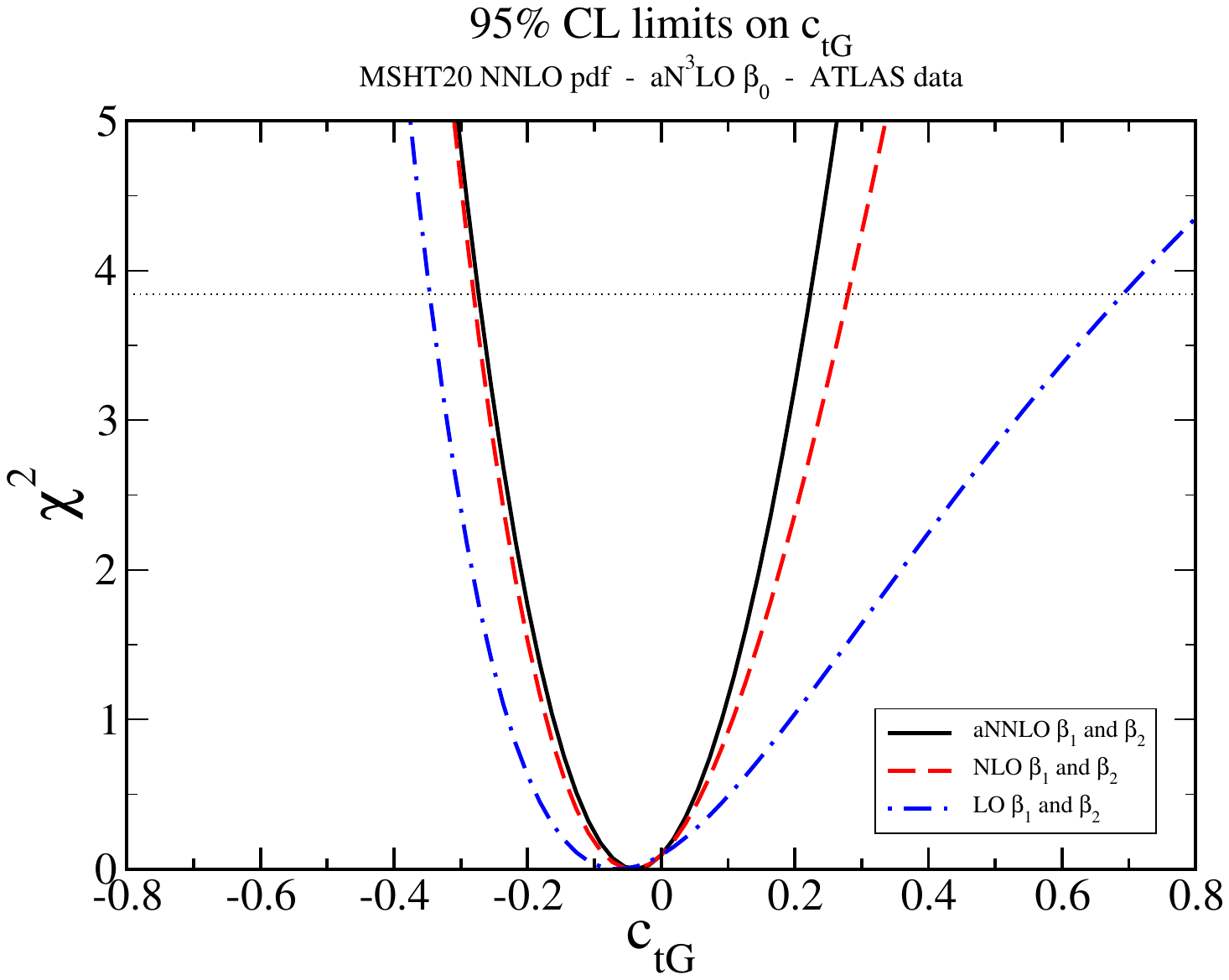}
\caption{Chi-squared as function of the effective operator coefficient $c_{tG}$ constructed using LO (blue dot-dashed curve), NLO (red dashed curve), and aNNLO (solid curve) SMEFT contributions $\beta_1$ and $\beta_2$. The left (right) plot uses $\beta_0$ computed at NNLO (aN$^3$LO) in QCD. The ATLAS measurement~\cite{ATLAS:2023gsl} is used as experimental input.}
\label{chi2atlas}
\end{center}
\end{figure}
\begin{figure}[htbp]
\begin{center}
\includegraphics[width=88mm]{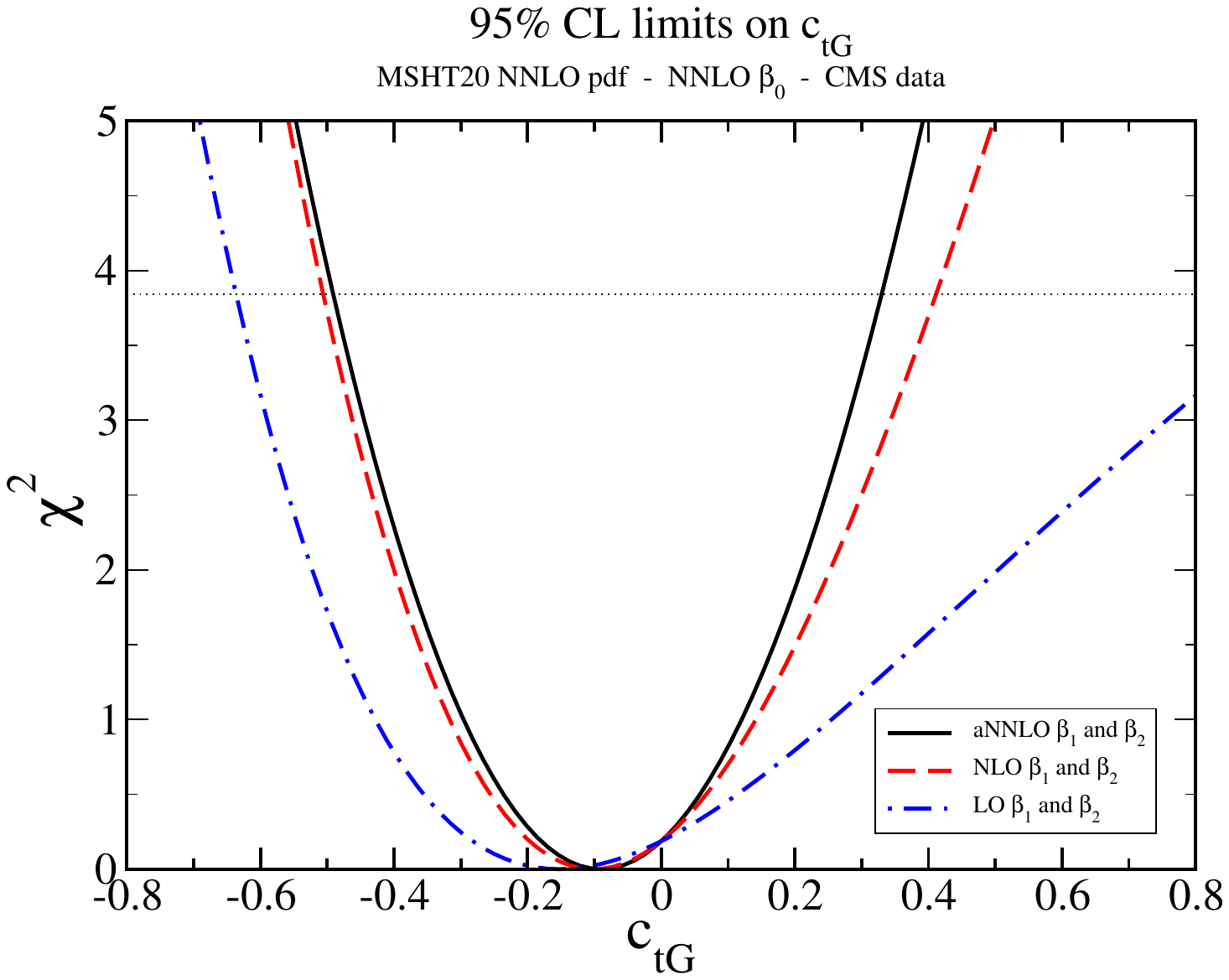}
\includegraphics[width=88mm]{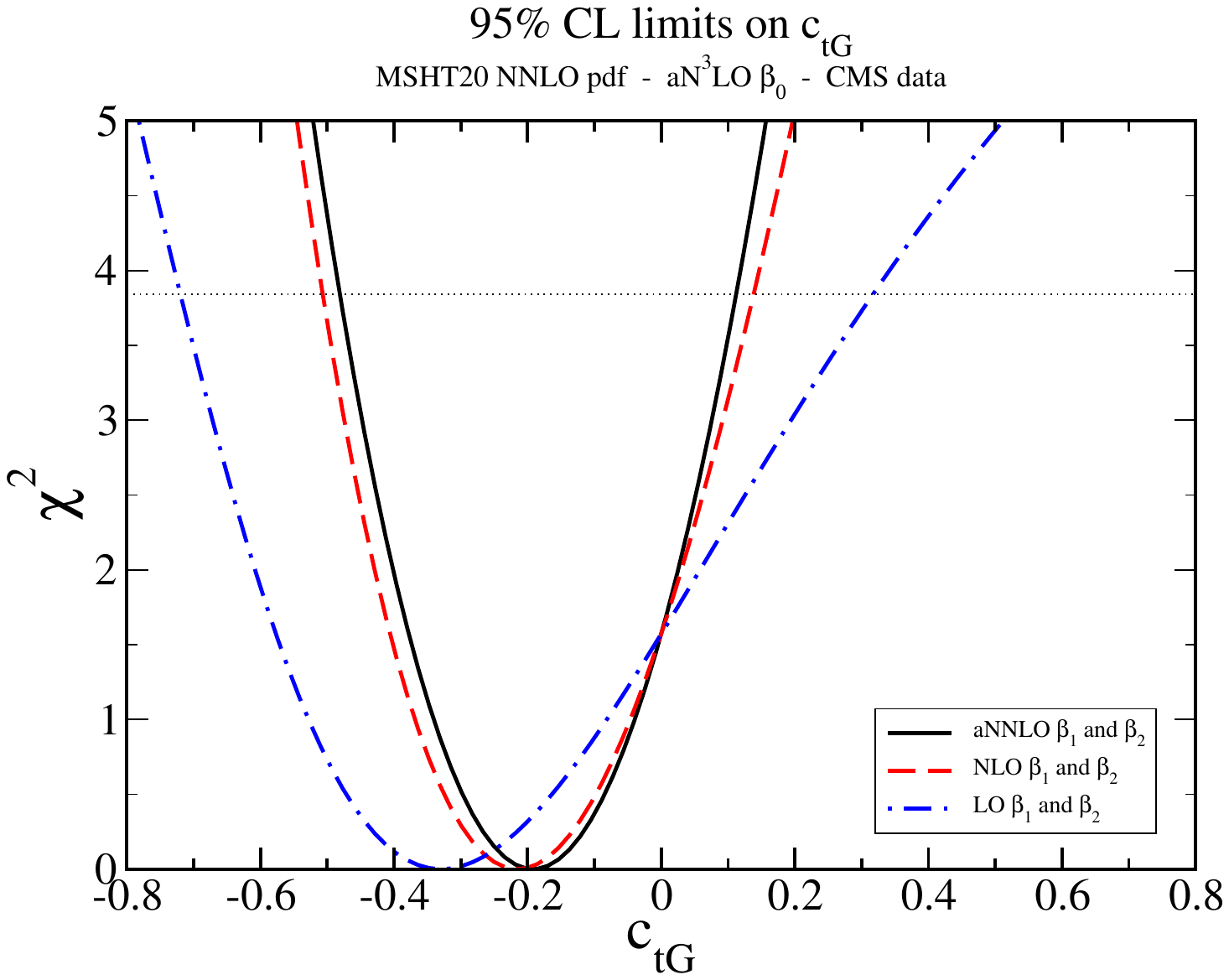}
\caption{Chi-squared as function of the effective operator coefficient $c_{tG}$ constructed using LO (blue dot-dashed curve), NLO (red dashed curve), and aNNLO (solid curve) SMEFT contributions $\beta_1$ and $\beta_2$ . The left (right) plot uses $\beta_0$ computed at NNLO (aN$^3$LO) in QCD. The CMS measurement~\cite{CMS:2021vhb} is used as experimental input.}
\label{chi2cms}
\end{center}
\end{figure}

In Fig.~\ref{chi2atlas} and Fig.~\ref{chi2cms} we show three curves representing the chi-squared function of Eq.~(\ref{chi2f}) obtained by using LO (blue dot-dashed line), NLO (red dashed line) and aNNLO (black solid line) QCD predictions for the SMEFT contributions $\beta_1$ and $\beta_2$. The chi-squared is a function of the effective operator coefficient and values of $c_{tG}$ are considered to be excluded at 95\% CL if $\chi^2(c_{tG})>3.841$. In both figures, the left plots use as SM contribution $\beta_0$ the NNLO QCD result taken from Table 1, while the right plots use the aN$^3$LO QCD result for $\beta_0$ taken from~\cite{KGT}. In Fig.~\ref{chi2atlas} we use as experimental input the ATLAS measurement of $829\pm 15$ pb~\cite{ATLAS:2023gsl}, while in Fig.~\ref{chi2cms} we use the CMS result of $791 \pm 25$ pb~\cite{CMS:2021vhb}.

Looking at Fig.~\ref{chi2atlas} we can see that, in both plots, the allowed region for the coefficient $c_{tG}$ is substantially reduced going from LO to NLO and it is further improved when going from NLO to aNNLO. For the left plot we have that the negative limit value reduces by about 2\% and the positive limit value reduces by about 25\% when considering aNNLO SMEFT contributions. For the right plot we have that the aNNLO contributions make the negative limit value to reduce by about 3\% and the positive limit value to reduce by about 25\%.

In analogy, the plots of Fig.~\ref{chi2cms} show that the allowed region for the coefficient $c_{tG}$ is also further improved when going from NLO to aNNLO. For the left plot we have that the negative limit value reduces by about 3\% and the positive limit value reduces by about 35\% when considering aNNLO SMEFT contributions. For the right plot we have that the aNNLO contributions make the negative limit value to reduce by about 5\% and the positive limit value to reduce by about 23\%.

\mysection{Top-quark $p_T$ distributions at 13 and 13.6 TeV}

As the resummation formalism in Section 2 indicates, soft-gluon resummation is derived not only for total cross sections but also for differential distributions. In this section we present results for the top-quark $p_T$ distribution for $t{\bar t}$ production at the LHC, including the effect of the dimension-six operator in Eq.~(\ref{opdef}) at 13 TeV and 13.6 TeV center-of-mass energies. For the binning of the $p_T$ distribution, we use the one of Refs.~\cite{CMS:2018adi,ATLAS:2019hxz} (six $p_T$ bins) which describes a top-quark transverse-momentum spectrum up to 550 GeV. In analogy to the total cross section case, we consider diagrams containing at most one effective operator insertion. Also in this case, the differential cross section can be written as a polynomial of second degree in the Wilson coefficient, 
\begin{equation}\label{dsigma_pt_smeft}
\frac{d\sigma(c_{tG})}{dp_T}=\frac{d\beta_0}{dp_T} + \frac{c_{tG}}{(\Lambda/1 {\rm TeV})^2} \frac{d\beta_1}{dp_T}
+\frac{c_{tG}^2}{(\Lambda/1 {\rm TeV})^4} \frac{d\beta_2}{dp_T} \, ,
\end{equation}
where $d\beta_0/dp_T$ is the SM differential cross section, $d\beta_1/dp_T$ is the contribution derived from the interference of SM diagrams and diagrams with one SMEFT insertion, while $d\beta_2/dp_T$ corresponds to the  terms derived by squaring diagrams with one SMEFT insertion. 

The SM contribution, $d\beta_0/dp_T$, is known up to aN$^3$LO in QCD \cite{NKan3lo1,NKan3lo2} (also including EW corrections \cite{KGT}) and here we use the NNLO QCD result.  As before, in the case of SMEFT contributions, the complete LO and NLO QCD results are calculated using {\small \sc MadGraph5\_aMC@NLO} and the {\small \sc SMEFT@NLO} model. Then, aNNLO distributions are obtained by adding second-order soft-gluon corrections to the complete NLO QCD result. We use MSHT20 NNLO pdf in our calculations and we set the factorization and renormalization scales equal to each other, with this common scale denoted by $\mu$. Here, the central results are obtained by setting $\mu=m_T$, where $m_T=(p_T^2+m_t^2)^{1/2}$ is the top-quark transverse mass. Scale uncertainties are obtained by varying the common scale $\mu$ in the range $m_T/2\leq \mu\leq 2 m_T$, and pdf uncertainties are also computed.

\begin{figure}[htbp]
\begin{center}
\includegraphics[width=88mm]{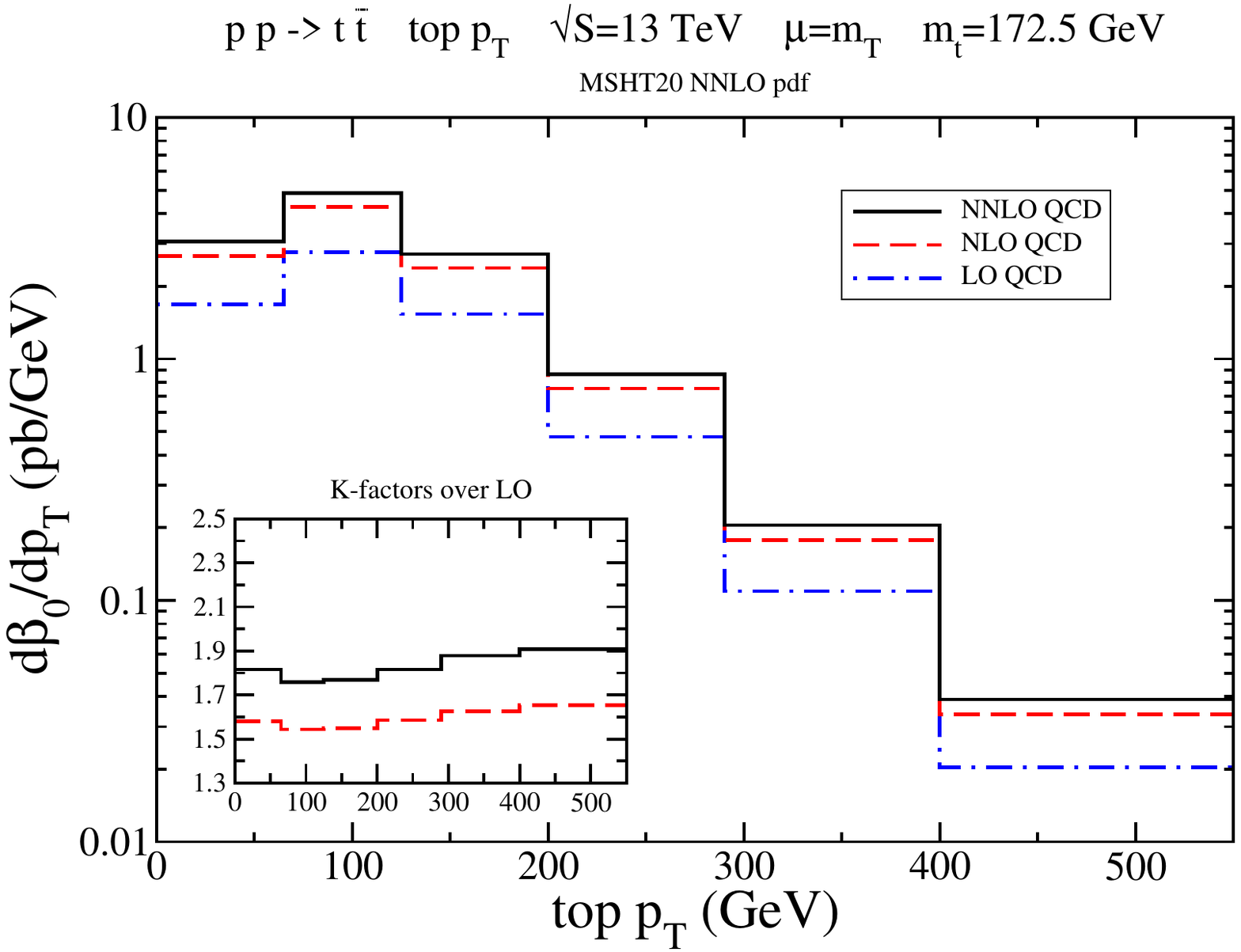}
\includegraphics[width=88mm]{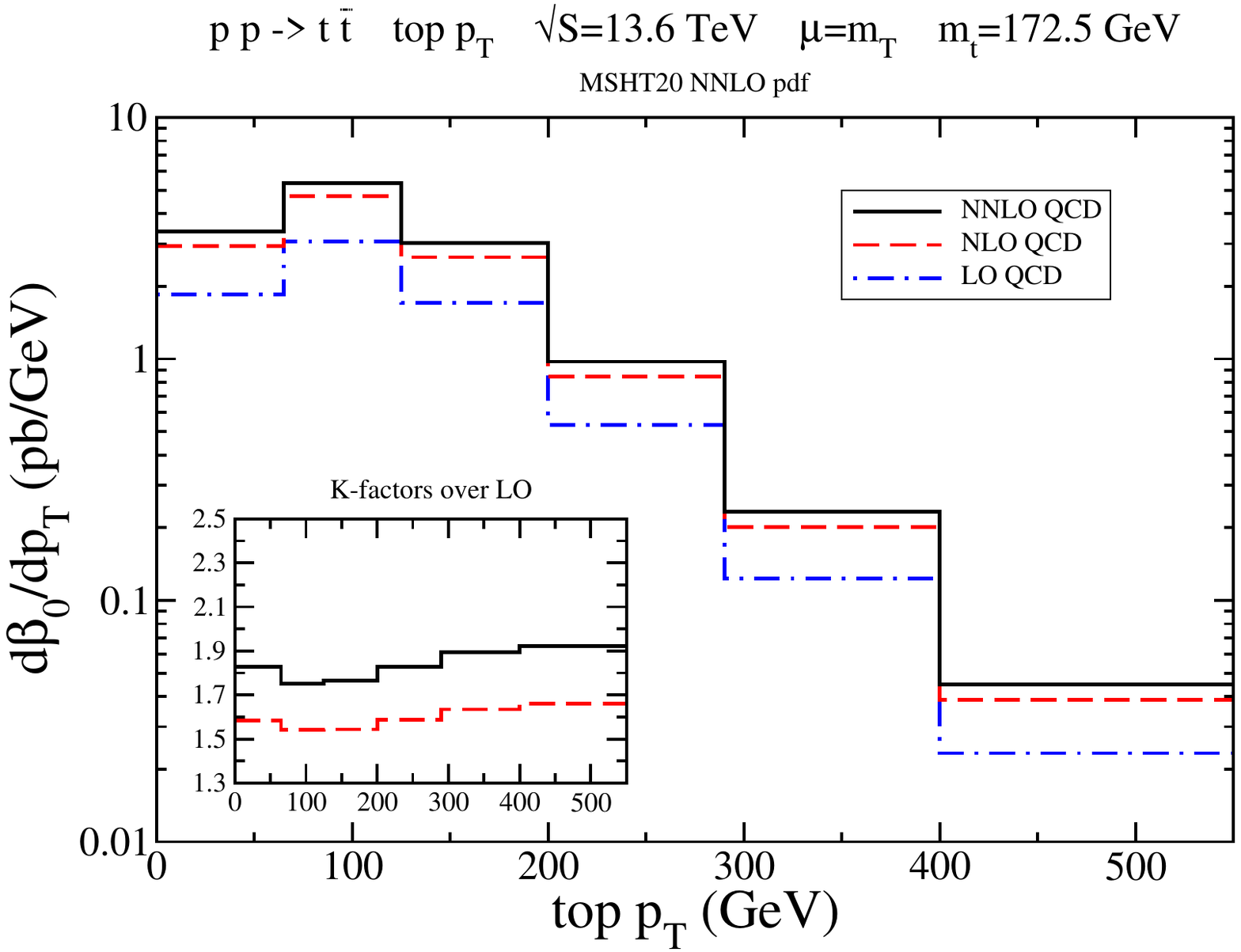}
\caption{SM contribution, $d\beta_0/dp_T$, to the top-quark $p_T$ distribution at different QCD perturbative orders (LO, NLO, and NNLO) with MSHT20 NNLO pdf  and $\mu = m_T$ computed for 13 TeV (left plot) and 13.6 TeV (right plot) center-of-mass energy. The inset plots represent $K$-factors over the LO QCD results.}
\label{ptb0plot}
\end{center}
\end{figure}

In Table 3 and Table 4 we present numerical results for the SM contribution, $d\beta_0/dp_T$, to the top-quark $p_T$ distribution in six $p_T$ bins at LO, NLO, and NNLO QCD using MSHT20 NNLO pdf; these results are computed for 13 TeV and 13.6 TeV center-of-mass energy, respectively. The central result in each bin is calculated at scale $\mu=m_T$ and reported together with scale variation $m_T/2\leq \mu\leq 2 m_T$ and pdf uncertainties. NNLO/NLO $K$-factors are also presented. Scale uncertainties at NLO are about $+13\%$ $-12\%$ for the first bin and $+15\%$ $-14\%$ for the last bin. At NNLO they improve substantially, being about $+5\%$ $-6\%$ for the first bin and $+5\%$ $-7\%$ for the last bin.
In Fig.~\ref{ptb0plot} we show the theoretical predictions (central value) for $d\beta_0/dp_T$ at $\sqrt{S} = 13$ TeV (left plot) and $\sqrt{S} = 13.6$ TeV (right plot) collision energy. $K$-factors over the LO QCD results are shown in each inset plot. We note that, in each plot, both the NLO/LO and the aNNLO/LO $K$-factors first decrease and then increase as the top-quark transverse momentum increases.

\begin{table}[htbp]
\begin{center}
\begin{tabular}{|c|c|c|c|c|c|} \hline
\multicolumn{5}{|c|}{SM contribution to the top-quark $p_T$ distribution at 13 TeV} \\ \hline
 $d\beta_0/dp_T$ (pb/GeV)&  LO &  NLO &  NNLO & NNLO/NLO\\ \hline
$0<p_T<65$ GeV & $1.69^{+0.45}_{-0.34}{}^{+0.02}_{-0.02} $ & $2.66^{+0.34}_{-0.31}{}^{+0.05}_{-0.03} $& $3.06^{+0.14}_{-0.19}{}^{+0.05}_{-0.04} $& 1.15\\ \hline
$65<p_T<125$ GeV & $2.77^{+0.76}_{-0.57}{}^{+0.05}_{-0.04} $ & $4.27^{+0.51}_{-0.50}{}^{+0.08}_{-0.06} $& $4.87^{+0.22}_{-0.29}{}^{+0.09}_{-0.07} $ & 1.14\\ \hline
$125<p_T<200$ GeV & $1.54^{+0.44}_{-0.33}{}^{+0.03}_{-0.02} $ & $2.38^{+0.29}_{-0.29}{}^{+0.05}_{-0.03} $& $2.71^{+0.13}_{-0.16}{}^{+0.06}_{-0.03} $ & 1.14\\ \hline
$200<p_T<290$ GeV & $0.476^{+0.143}_{-0.103}{}^{+0.011}_{-0.007} $&$0.755^{+0.100}_{-0.096}{}^{+0.019}_{-0.011} $& $0.864^{+0.040}_{-0.058}{}^{+0.021}_{-0.013} $& 1.15\\ \hline
$290<p_T<400$ GeV & $0.109^{+0.034}_{-0.024}{}^{+0.003}_{-0.002} $& $0.177^{+0.026}_{-0.024}{}^{+0.005}_{-0.003} $& $0.205^{+0.010}_{-0.014}{}^{+0.006}_{-0.004} $& 1.16\\ \hline
$400<p_T<550$ GeV & $0.0204^{+0.0065}_{-0.0047}{}^{+0.0006}_{-0.0004} $& $0.0336^{+0.0051}_{-0.0047}{}^{+0.0011}_{-0.0006} $& $0.0389^{+0.0019}_{-0.0027}{}^{+0.0013}_{-0.0008} $& 1.16\\ \hline
\end{tabular}
\caption[]{SM contribution, $d\beta_0/dp_T$, to the top-quark $p_T$ distribution at LO, NLO, and NNLO at the LHC with $\sqrt{S}=13$ TeV, $m_t=172.5$ GeV, and MSHT20 NNLO pdf. The central results are for $\mu=m_T$ and are shown together with scale and pdf uncertainties. The NNLO/NLO $K$-factors are also shown.}
\label{tableptb0}
\end{center}
\end{table}

\begin{table}[htbp]
\begin{center}
\begin{tabular}{|c|c|c|c|c|} \hline
\multicolumn{5}{|c|}{SM contribution to the top-quark $p_T$ distribution at 13.6 TeV} \\ \hline
 $d\beta_0/dp_T$ (pb/GeV)&  LO &  NLO &  NNLO& NNLO/NLO\\ \hline
$0<p_T<65$ GeV & $1.85^{+0.50}_{-0.36}{}^{+0.03}_{-0.02} $ & $2.93^{+0.37}_{-0.34}{}^{+0.05}_{-0.03} $& $3.38^{+0.16}_{-0.21}{}^{+0.06}_{-0.04}  $& 1.15 \\ \hline
$65<p_T<125$ GeV & $3.06^{+0.83}_{-0.62}{}^{+0.04}_{-0.04} $ & $4.72^{+0.57}_{-0.54}{}^{+0.09}_{-0.06} $& $5.36^{+0.24}_{-0.31}{}^{+0.10}_{-0.06}  $ & 1.14\\ \hline
$125<p_T<200$ GeV & $1.71^{+0.48}_{-0.36}{}^{+0.03}_{-0.03} $ & $2.64^{+0.32}_{-0.31}{}^{+0.06}_{-0.03} $& $3.02^{+0.13}_{-0.19}{}^{+0.06}_{-0.05}  $ & 1.14 \\ \hline
$200<p_T<290$ GeV & $0.533^{+0.158}_{-0.114}{}^{+0.013}_{-0.007} $&$0.846^{+0.112}_{-0.107}{}^{+0.021}_{-0.012} $& $0.974^{+0.047}_{-0.065}{}^{+0.025}_{-0.015}  $& 1.15 \\ \hline
$290<p_T<400$ GeV & $0.123^{+0.039}_{-0.027}{}^{+0.004}_{-0.002} $& $0.201^{+0.029}_{-0.027}{}^{+0.006}_{-0.003} $& $0.233^{+0.011}_{-0.017}{}^{+0.006}_{-0.004}  $& 1.16 \\ \hline
$400<p_T<550$ GeV & $0.0233^{+0.0074}_{-0.0053}{}^{+0.0007}_{-0.0004} $& $0.0387^{+0.0059}_{-0.0054}{}^{+0.0013}_{-0.0007} $& $0.0448^{+0.0021}_{-0.0032}{}^{+0.0014}_{-0.0009}  $ & 1.16\\ \hline
\end{tabular}
\caption[]{SM contribution, $d\beta_0/dp_T$, to the top-quark $p_T$ distribution at LO, NLO, and NNLO at the LHC with $\sqrt{S}=13.6$ TeV, $m_t=172.5$ GeV, and MSHT20 NNLO pdf. The central results are for $\mu=m_T$ and are shown together with scale and pdf uncertainties. The NNLO/NLO $K$-factors are also shown.}
\label{table2ptb0}
\end{center}
\end{table}

\begin{figure}[htbp]
\begin{center}
\includegraphics[width=88mm]{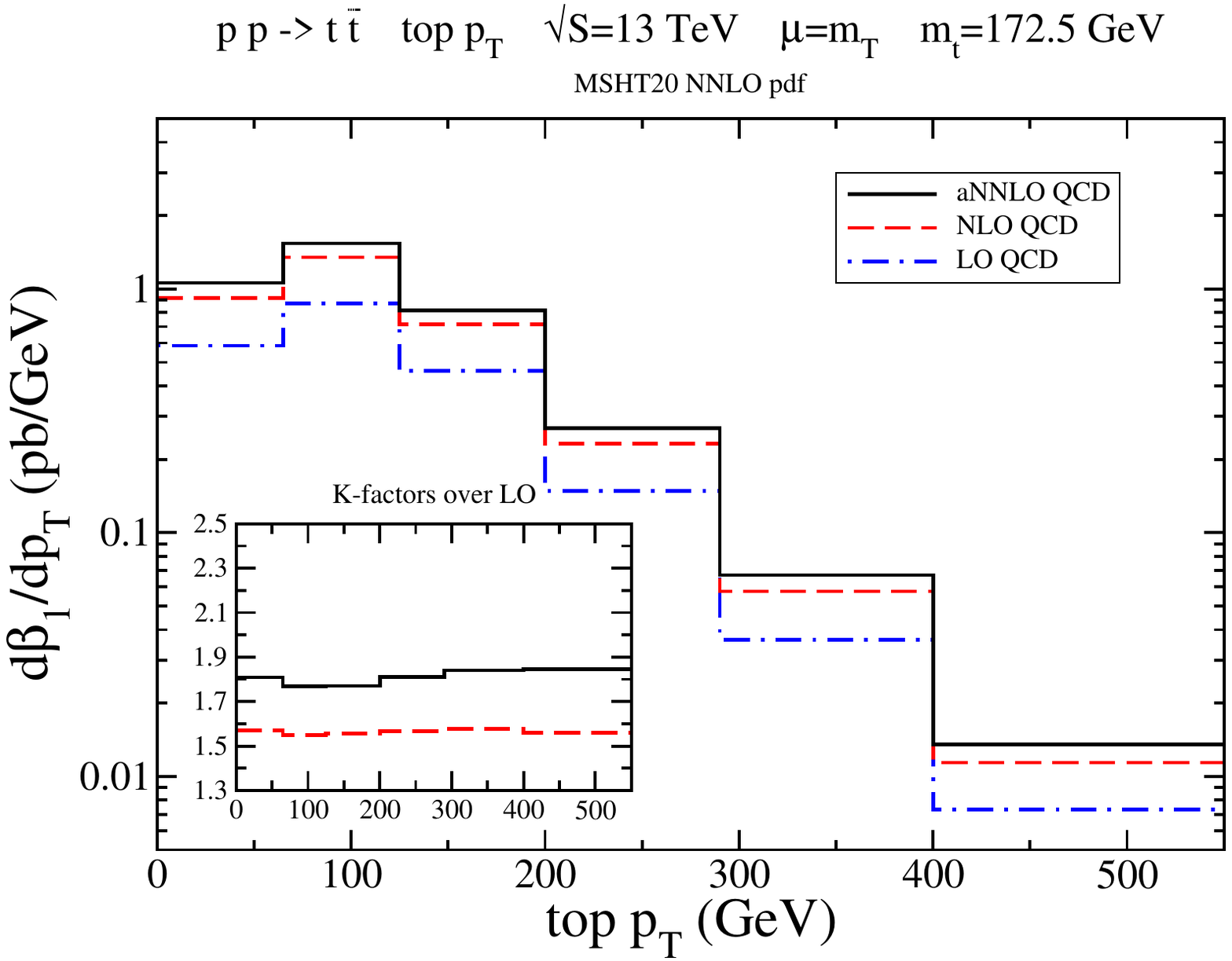}
\includegraphics[width=88mm]{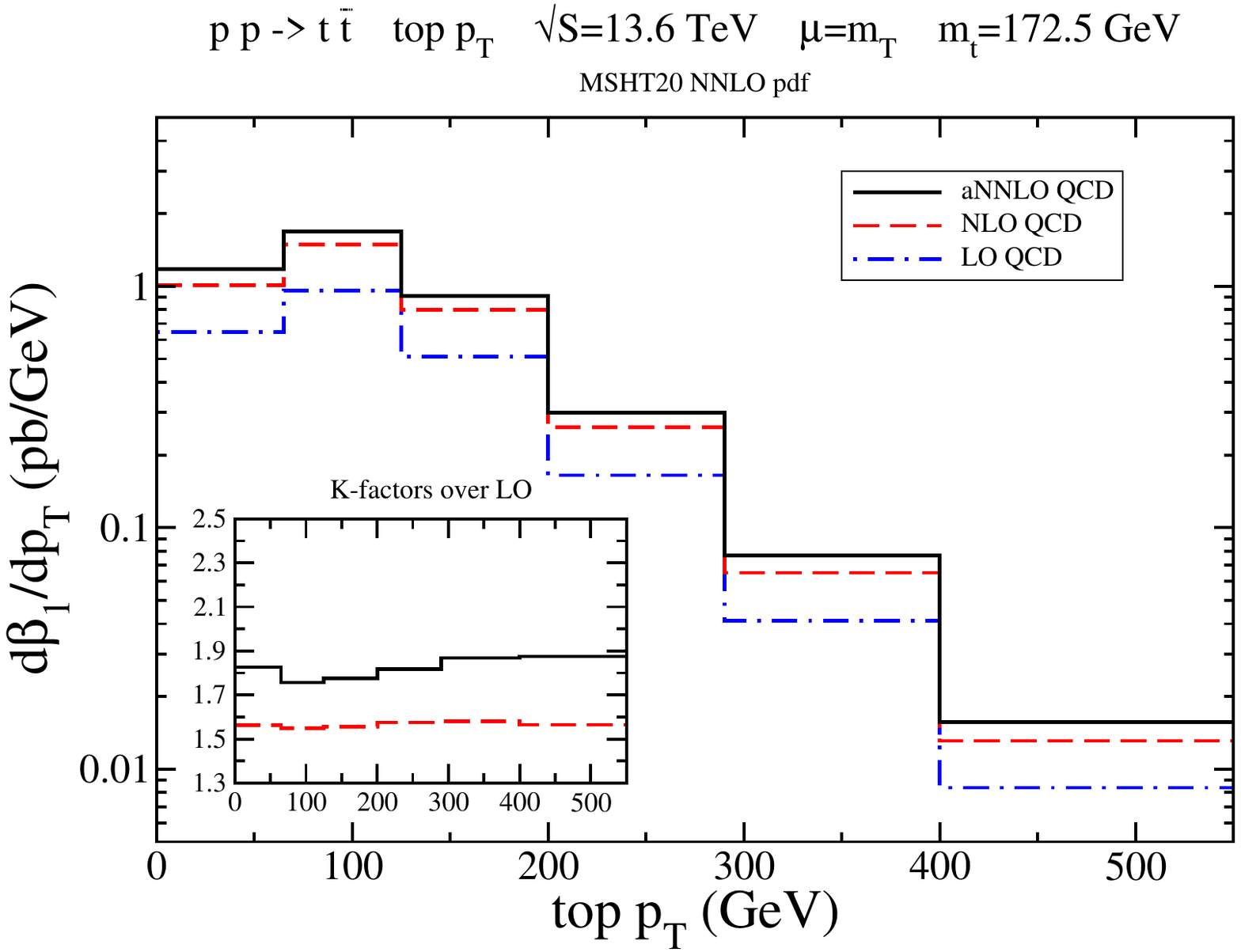}
\caption{SMEFT linear contribution, $d\beta_1/dp_T$, to the top-quark $p_T$ distribution at different QCD perturbative orders (LO, NLO, and aNNLO) with MSHT20 NNLO pdf  and $\mu = m_T$ computed for 13 TeV (left plot) and 13.6 TeV (right plot) center-of-mass energy. The inset plots represent $K$-factors over the LO QCD results.}
\label{ptb1plot}
\end{center}
\end{figure}

In Table 5 and Table 6 we present numerical results for the SMEFT linear contribution, $d\beta_1/dp_T$, to the top-quark $p_T$ distribution in six $p_T$ bins at LO, NLO, and aNNLO QCD using MSHT20 NNLO pdf; these results are computed for 13 TeV and 13.6 TeV center-of-mass energy, respectively. The central result in each bin is calculated at scale $\mu=m_T$ and reported together with scale variation $m_T/2\leq \mu\leq 2 m_T$ and pdf uncertainties. aNNLO/NLO $K$-factors are also presented. Scale uncertainties are very similar to the $d\beta_0/dp_T$ case: at NLO they are about $+12\%$ $-12\%$ for the first bin and $+13\%$ $-13\%$ for the last bin. At aNNLO they improve substantially, being about $+5\%$ $-7\%$ for the first bin and $+5\%$ $-7\%$ for the last bin.
In Fig.~\ref{ptb1plot} we show the theoretical predictions (central value) for $d\beta_1/dp_T$ at $\sqrt{S} = 13$ TeV (left plot) and $\sqrt{S} = 13.6$ TeV (right plot) collision energy. $K$-factors over the LO QCD results are shown in each inset plot. Here we note that, in each plot, the NLO/LO $K$-factor behavior presents a more flat dependence on the top-quark transverse momentum, while the aNNLO/LO $K$-factor behavior is similar, although less pronounced, to the SM case.

\begin{table}[htbp]
\begin{center}
\begin{tabular}{|c|c|c|c|c|} \hline
\multicolumn{5}{|c|}{SMEFT linear contribution to the top-quark $p_T$ distribution at 13 TeV} \\ \hline
 $d\beta_1/dp_T$ (pb/GeV)& LO &  NLO &  aNNLO& aNNLO/NLO\\ \hline
$0<p_T<65$ GeV & $0.586^{+0.159}_{-0.117}{}^{+0.011}_{-0.007} $ & $0.920^{+0.112}_{-0.108}{}^{+0.016}_{-0.012} $& $1.06^{+0.05}_{-0.07}{}^{+0.02}_{-0.01} $&1.15\\ \hline
$65<p_T<125$ GeV & $0.871^{+0.240}_{-0.178}{}^{+0.016}_{-0.012} $ & $1.35^{+0.16}_{-0.16}{}^{+0.03}_{-0.02} $& $1.54^{+0.07}_{-0.09}{}^{+0.03}_{-0.02} $ &1.14\\ \hline
$125<p_T<200$ GeV & $0.461^{+0.132}_{-0.096}{}^{+0.010}_{-0.006} $ & $0.717^{+0.088}_{-0.087}{}^{+0.015}_{-0.010} $& $0.816^{+0.039}_{-0.048}{}^{+0.017}_{-0.011} $&1.14 \\ \hline
$200<p_T<290$ GeV & $0.148^{+0.044}_{-0.032}{}^{+0.003}_{-0.002} $&$0.232^{+0.030}_{-0.029} {}^{+0.005}_{-0.004}$& $0.268^{+0.012}_{-0.018}{}^{+0.006}_{-0.004} $&1.16\\ \hline
$290<p_T<400$ GeV & $0.0364^{+0.0113}_{-0.0080}{}^{+0.0010}_{-0.0005} $& $0.0574^{+0.0076}_{-0.0074}{}^{+0.0016}_{-0.009} $& $0.0670^{+0.0033}_{-0.0046}{}^{+0.0018}_{-0.0011} $&1.17\\ \hline
$400<p_T<550$ GeV & $0.00731^{+0.00233}_{-0.00165} {}^{+0.00021}_{-0.00013}$& $0.0114^{+0.0015}_{-0.0015}{}^{+0.0004}_{-0.0002} $& $0.0135^{+0.0007}_{-0.0009}{}^{+0.0004}_{-0.0003} $&1.18\\ \hline
\end{tabular}
\caption[]{SMEFT linear contribution, $d\beta_1/dp_T$, to the top-quark $p_T$ distribution at LO, NLO, and aNNLO at the LHC with $\sqrt{S}=13$ TeV, $m_t=172.5$ GeV, and MSHT20 NNLO pdf. The central results are for $\mu=m_T$ and are shown together with scale and pdf uncertainties. The aNNLO/NLO $K$-factors are also shown.}
\label{tableptb1}
\end{center}
\end{table}

\begin{table}[htbp]
\begin{center}
\begin{tabular}{|c|c|c|c|c|} \hline
\multicolumn{5}{|c|}{SMEFT linear contribution to the top-quark $p_T$ distribution at 13.6 TeV} \\ \hline
 $d\beta_1/dp_T$ (pb/GeV)&  LO &  NLO &  aNNLO& aNNLO/NLO\\ \hline
$0<p_T<65$ GeV & $0.646^{+0.171}_{-0.128}{}^{+0.011}_{-0.008} $ & $1.01^{+0.12}_{-0.12}{}^{+0.02}_{-0.01} $& $1.18^{+0.06}_{-0.07}{}^{+0.02}_{-0.01} $&1.17\\ \hline
$65<p_T<125$ GeV & $0.962^{+0.263}_{-0.194}{}^{+0.017}_{-0.012} $ & $1.49^{+0.18}_{-0.17}{}^{+0.03}_{-0.02} $& $1.69^{+0.08}_{-0.10}{}^{+0.03}_{-0.02} $ &1.14\\ \hline
$125<p_T<200$ GeV & $0.512^{+0.144}_{-0.106}{}^{+0.010}_{-0.007} $ & $0.797^{+0.098}_{-0.096}{}^{+0.016}_{-0.011} $& $0.909^{+0.039}_{-0.057}{}^{+0.018}_{-0.013} $&1.14\\ \hline
$200<p_T<290$ GeV & $0.165^{+0.049}_{-0.035}{}^{+0.004}_{-0.002} $&$0.260^{+0.034}_{-0.032}{}^{+0.006}_{-0.004} $& $0.300^{+0.015}_{-0.020}{}^{+0.007}_{-0.005} $ &1.15\\ \hline
$290<p_T<400$ GeV & $0.0411^{+0.0126}_{-0.0090}{}^{+0.0011}_{-0.0006} $& $0.0650^{+0.0087}_{-0.0084} {}^{+0.0017}_{-0.0011}$& $0.0768^{+0.0036}_{-0.0056}{}^{+0.0020}_{-0.0013} $&1.18\\ \hline
$400<p_T<550$ GeV & $0.00837^{+0.00265}_{-0.00187}{}^{+0.00024}_{-0.00014} $& $0.0131^{+0.0018}_{-0.0017}{}^{+0.0004}_{-0.0002} $& $0.0157^{+0.0007}_{-0.0011}{}^{+0.0005}_{-0.0002} $&1.20\\ \hline
\end{tabular}
\caption[]{SMEFT linear contribution, $d\beta_1/dp_T$, to the top-quark $p_T$ distribution at LO, NLO, and aNNLO at the LHC with $\sqrt{S}=13.6$ TeV, $m_t=172.5$ GeV, and MSHT20 NNLO pdf. The central results are for $\mu=m_T$ and are shown together with scale and pdf uncertainties. The aNNLO/NLO $K$-factors are also shown.}
\label{tablep2tb1}
\end{center}
\end{table}

\begin{figure}[htbp]
\begin{center}
\includegraphics[width=88mm]{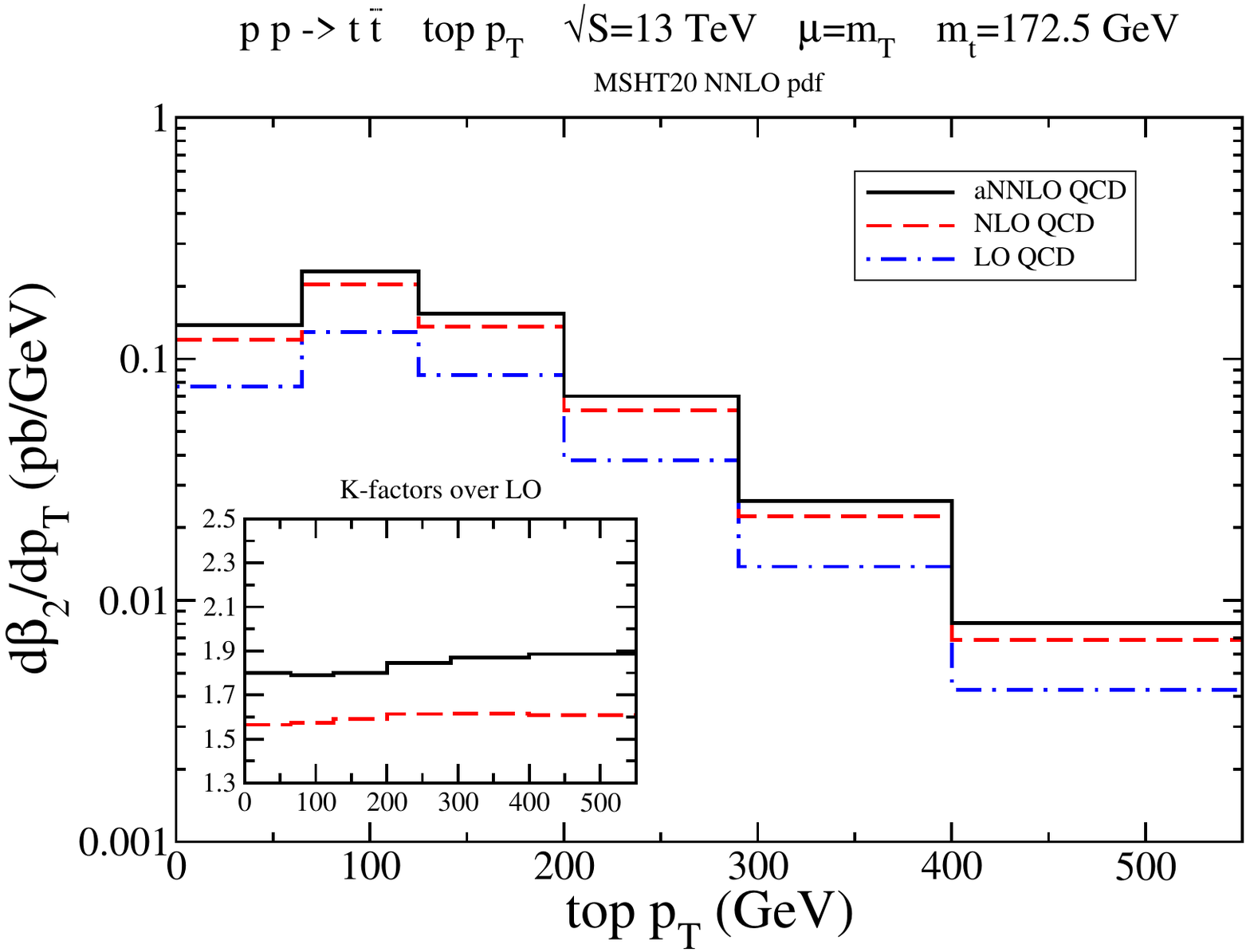}
\includegraphics[width=88mm]{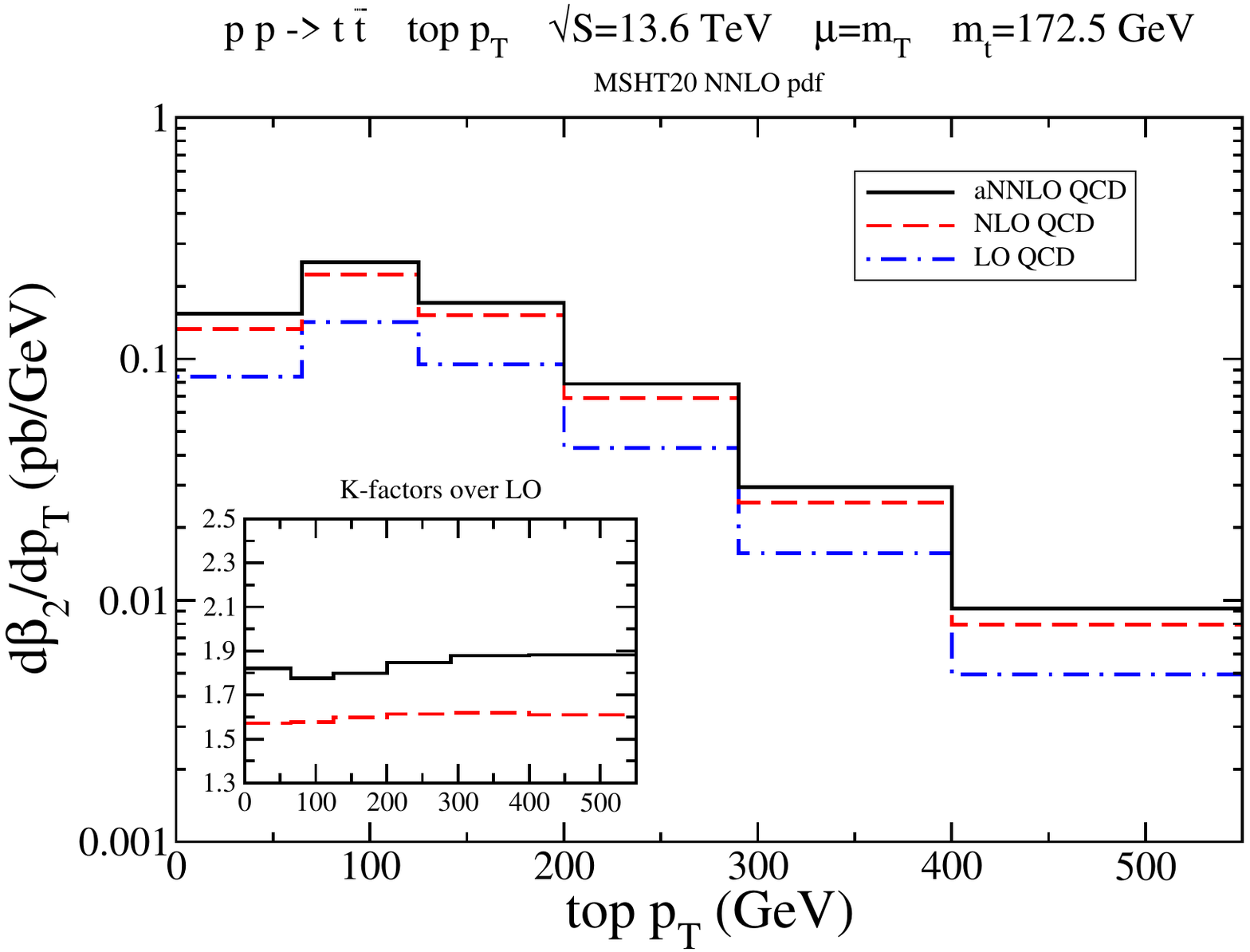}
\caption{SMEFT quadratic contribution, $d\beta_2/dp_T$, to the top-quark $p_T$ distribution at different QCD perturbative orders (LO, NLO, and aNNLO) with MSHT20 NNLO pdf  and $\mu = m_T$ computed for 13 TeV (left plot) and 13.6 TeV (right plot) center-of-mass energy. The inset plots represent $K$-factors over the LO QCD results.}
\label{ptb2plot}
\end{center}
\end{figure}

In Table 7 and Table 8 we present numerical results for the SMEFT quadratic contribution, $d\beta_2/dp_T$, to the top-quark $p_T$ distribution in six $p_T$ bins at LO, NLO, and aNNLO QCD using MSHT20 NNLO pdf; these results are computed for 13 TeV and 13.6 TeV center of mass energy, respectively. The central result in each bin is calculated at scale $\mu=m_T$ and reported together with scale variation $m_T/2\leq \mu\leq 2 m_T$ and pdf uncertainties. aNNLO/NLO $K$-factors are also presented. Also here, scale uncertainties are very similar to the SM and $d\beta_1/dp_T$ case: at NLO they are about $+13\%$ $-12\%$ for the first bin and $+14\%$ $-14\%$ for the last bin. At aNNLO they improve substantially, being about $+4\%$ $-7\%$ for the first bin and $+5\%$ $-7\%$ for the last bin.
In Fig.~\ref{ptb2plot} we show the theoretical predictions (central value) for $d\beta_2/dp_T$ at $\sqrt{S} = 13$ TeV (left plot) and $\sqrt{S} = 13.6$ TeV (right plot) collision energy. $K$-factors over the LO QCD results are shown in each inset plot. The behavior of the $K$-factors as function of top-quark transverse momentum is similar to the previous case of $d\beta_1/dp_T$.

In the case of the $p_T$ distribution, the binned SMEFT and SM $K$-factors might present sizable differences. For instance, at 13 TeV, the NLO over LO $K$-factors of $d\beta_1/dp_T$ differ by 0.4-5.3\% with respect to the SM, depending on the bin. In analogy, the NLO over LO $K$-factors of $d\beta_2/dp_T$ differ by 0.1-2.8\% with respect to the SM, depending on the bin.  Again at 13 TeV, the aNNLO over LO $K$-factors of $d\beta_1/dp_T$ differ by 0.1-3.2\% with respect to the SM, depending on the bin, while, the aNNLO over LO $K$-factors of $d\beta_2/dp_T$ differ by 0.6-2.0\% with respect to the SM, depending on the bin. 

\begin{table}[htbp]
\begin{center}
\begin{tabular}{|c|c|c|c|c|} \hline
\multicolumn{5}{|c|}{SMEFT quadratic contribution to the top-quark $p_T$ distribution at 13 TeV} \\ \hline
$d\beta_2/dp_T$ (pb/GeV) & LO &  NLO &  aNNLO & \!\!\!\! aNNLO/NLO \!\!\!\!\! \\ \hline
$0<p_T<65$ GeV & $0.0767^{+0.0209}_{-0.0154}{}^{+0.0015}_{-0.0010} $ & $0.120^{+0.015}_{-0.014} {}^{+0.003}_{-0.001}$& $0.138^{+0.006}_{-0.009}{}^{+0.002}_{-0.002}$&1.15 \\ \hline
$65<p_T<125$ GeV & $0.129^{+0.036}_{-0.027}{}^{+0.002}_{-0.002} $ & $0.203^{+0.025}_{-0.025}{}^{+0.004}_{-0.003} $& $0.231^{+0.010}_{-0.014}{}^{+0.005}_{-0.003} $&1.14\\ \hline
$125<p_T<200$ GeV & $0.0855^{+0.0248}_{-0.0181}{}^{+0.0019}_{-0.0013} $ & $0.136^{+0.018}_{-0.017}{}^{+0.003}_{-0.002} $& $0.154^{+0.007}_{-0.009}{}^{+0.004}_{-0.002} $&1.13 \\ \hline
$200<p_T<290$ GeV & $0.0380^{+0.0116}_{-0.0083}{}^{+0.0010}_{-0.0006} $&$0.0613^{+0.0083}_{-0.0080}{}^{+0.0016}_{-0.0011} $& $0.0701^{+0.0032}_{-0.0047}{}^{+0.0019}_{-0.0012} $&1.14\\ \hline
$290<p_T<400$ GeV & $0.0138^{+0.0044}_{-0.0031}{}^{+0.0004}_{-0.0002} $& $0.0223^{+0.0031}_{-0.0030}{}^{+0.0007}_{-0.0004} $& $0.0258^{+0.0013}_{-0.0018}{}^{+0.0008}_{-0.0005} $&1.16\\ \hline
$400<p_T<550$ GeV & $0.00427^{+0.00141}_{-0.00100}{}^{+0.00015}_{-0.00010} $& $0.00687^{+0.00095}_{-0.00094}{}^{+0.00025}_{-0.00016} $& $0.00805^{+0.00039}_{-0.00056}{}^{+0.00030}_{-0.00018} $&1.17\\ \hline
\end{tabular}
\caption[]{SMEFT quadratic contribution, $d\beta_2/dp_T$, to the top-quark $p_T$ distribution at LO, NLO, and aNNLO at the LHC with $\sqrt{S}=13$ TeV, $m_t=172.5$ GeV, and MSHT20 NNLO pdf. The central results are for $\mu=m_T$ and are shown together with scale and pdf uncertainties. The aNNLO/NLO $K$-factors are also shown.}
\label{tableptb2}
\end{center}
\end{table}

\begin{table}[htbp]
\begin{center}
\begin{tabular}{|c|c|c|c|c|} \hline
\multicolumn{5}{|c|}{SMEFT quadratic contribution to the top-quark  $p_T$ distribution at 13.6 TeV} \\ \hline
 $d\beta_2/dp_T$ (pb/GeV)&  LO &  NLO & aNNLO& \!\!\!\! aNNLO/NLO \!\!\!\!\! \\ \hline
$0<p_T<65$ GeV & $0.0846^{+0.0226}_{-0.0169}{}^{+0.0015}_{-0.0011} $ & $0.133^{+0.016}_{-0.016}{}^{+0.002}_{-0.002} $& $0.154^{+0.007}_{-0.009}{}^{+0.002}_{-0.002} $&1.16\\ \hline
$65<p_T<125$ GeV & $0.142^{+0.039}_{-0.029}{}^{+0.003}_{-0.002} $ & $0.224^{+0.028}_{-0.027}{}^{+0.004}_{-0.003} $& $0.252^{+0.011}_{-0.015}{}^{+0.005}_{-0.003} $ &1.13\\ \hline
$125<p_T<200$ GeV & $0.0951^{+0.0272}_{-0.0199}{}^{+0.0021}_{-0.0014} $ & $0.152^{+0.019}_{-0.019}{}^{+0.003}_{-0.003} $& $0.171^{+0.007}_{-0.011}{}^{+0.003}_{-0.003} $ &1.13\\ \hline
$200<p_T<290$ GeV & $0.0427^{+0.0128}_{-0.0093}{}^{+0.0011}_{-0.0007} $&$0.0689^{+0.0094}_{-0.0089}{}^{+0.0018}_{-0.0011} $& $0.0789^{+0.0038}_{-0.0053}{}^{+0.0021}_{-0.0013} $ &1.15\\ \hline
$290<p_T<400$ GeV & $0.0157^{+0.0049}_{-0.0035}{}^{+0.0005}_{-0.0003} $& $0.0254^{+0.0036}_{-0.0034}{}^{+0.0008}_{-0.0004} $& $0.0295^{+0.0014}_{-0.0022}{}^{+0.0009}_{-0.0005} $&1.16\\ \hline
$400<p_T<550$ GeV & $0.00492^{+0.00161}_{-0.00114}{}^{+0.00017}_{-0.00011} $& $0.00792^{+0.00109}_{-0.00107}{}^{+0.00029}_{-0.00017} $& $0.00926^{+0.00043}_{-0.00066}{}^{+0.00034}_{-0.00020} $&1.17\\ \hline
\end{tabular}
\caption[]{SMEFT quadratic contribution, $d\beta_2/dp_T$, to the top-quark $p_T$ distribution at LO, NLO, and aNNLO at the LHC with $\sqrt{S}=13.6$ TeV, $m_t=172.5$ GeV, and MSHT20 NNLO pdf. The central results are for $\mu=m_T$ and are shown together with scale and pdf uncertainties. The aNNLO/NLO $K$-factors are also shown.}
\label{table2ptb2}
\end{center}
\end{table}

\mysection{Conclusions}

In this paper we have provided a study of higher-order corrections to $t{\bar t}$ production at the LHC in the SMEFT framework. More specifically, we have added for the first time soft-gluon corrections at aNNLO to the complete QCD corrections at NLO, in the presence of the chromomagnetic dipole operator.  We have calculated SM and SMEFT contributions to the total cross sections at the LHC for two values of the collider energy, namely 13 TeV and 13.6 TeV. The SM and SMEFT contributions to the total cross sections get a similar large enhancement (50\%) from the complete NLO QCD corrections both at 13 TeV and 13.6 TeV center-of-mass energy. The additional aNNLO QCD corrections are also similar and significant for both SM and SMEFT contributions, accounting for another 17\% enhancement of the cross section, and  make the theoretical uncertainties from scale variation to get substantially reduced. In order to estimate the effects that our aNNLO QCD calculations might have on SMEFT global fits, we use our total cross section results at 13 TeV to set bounds on the size of the effective operator coefficient $c_{tG}$, comparing the theoretical predictions with the most recent ATLAS~\cite{ATLAS:2023gsl} and CMS~\cite{CMS:2021vhb} results. We found that aNNLO corrections improve the lower bound on the $c_{tG}$ coefficient by 2 to 5\%, while the upper bound reduces by 23 to 35\%, depending on the considered experimental input and SM prediction. Therefore, including these aNNLO contributions is crucial to improve further the sensitivity on SMEFT operators in global-fit analyses.

We have also computed SM and SMEFT contributions to the top-quark $p_T$ distribution up to aNNLO in QCD, considering six $p_T$ bins up to 550 GeV. These contributions receive large corrections at NLO and further significant enhancements at aNNLO, similar to the total cross section case. 
The top-quark transverse-momentum distribution has been used as observable in recent $t\bar t$ experimental analyses~\cite{CMS:2018adi,ATLAS:2022xfj,ATLAS:2022mlu} to set limits on the chromomagnetic dipole operator and other SMEFT operators. Therefore, our aNNLO theoretical calculation of the modifications induced by the chromomagnetic dipole operator to the top $p_T$ distribution is relevant and should be adopted when performing  fits in order to improve the sensitivity to this effective operator in future SMEFT searches.

\section*{Acknowledgements}
This material is based upon work supported by the National Science Foundation under Grant No. PHY 2112025.

\end{document}